\documentclass{jfm}
\usepackage{upmath}
\usepackage{amsmath}
\usepackage{amsfonts}
\usepackage{amssymb}
\usepackage{pgf,pgfarrows}
\usepackage{wasysym}
\usepackage{graphicx}
\usepackage[export]{adjustbox}
\usepackage{subfig}
\usepackage{natbib}
\usepackage{upgreek}
\usepackage{mathrsfs}
\usepackage{float}
\usepackage{caption}
\usepackage{soul}
\usepackage{textcomp}
\usepackage{stmaryrd}

\ifCUPmtlplainloaded \else
  \checkfont{eurm10}
  \iffontfound
    \IfFileExists{upmath.sty}
      {\typeout{^^JFound AMS Euler Roman fonts on the system,
                   using the 'upmath' package.^^J}
}
      {\typeout{^^JFound AMS Euler Roman fonts on the system, but you
                   dont seem to have the}
       \typeout{'upmath' package installed. JFM.cls can take advantage
                 of these fonts,^^Jif you use 'upmath' package.^^J}
       
      }
  \else
    
  \fi
\fi
\ifCUPmtlplainloaded \else
  \checkfont{msam10}
  \iffontfound
    \IfFileExists{amssymb.sty}
      {\typeout{^^JFound AMS Symbol fonts on the system, using the
                'amssymb' package.^^J}
  \let\leq=\leqslant
         \let\geq=\geqslant
      }{}
  \fi
\fi
\ifCUPmtlplainloaded \else
  \IfFileExists{amsbsy.sty}
    {\typeout{^^JFound the 'amsbsy' package on the system, using it.^^J}
}
    {\providecommand\boldsymbol[1]{\mbox{\boldmath $##1$}}}
\fi

\def\ee{{\rm e}}
\def\ii{{\rm i}}

\affiliation{
$^1$ Environmental and Geophysical Fluids Group, Department of Mechanical Engineering, Indian Institute of Technology, Kanpur, U.P. 208016, India.}

\pubyear{2012}
\volume{xxx}
\pagerange{xxx--yyy}
\title[On a new non-Boussinesq instability in stratified lakes and oceans]{On a new non-Boussinesq instability in stratified lakes and oceans}
\author[M.H.~Shete and A.~Guha]
{
M\ls I\ls H\ls I\ls R\ns H\ls.\ns  S\ls H\ls E\ls T\ls E\ls $^{1}$
\and
A\ls N\ls I\ls R\ls B\ls A\ls N\ns G\ls U\ls H\ls A$^{1}$\footnote{Electronic mail for correspondence: anirbanguha.ubc@gmail.com }
}
\date{?? and in revised form ??}

\begin{document}

\maketitle

\begin{abstract}
Lakes, as well as many other geophysical flows, are shallow, density stratified, and contain a free-surface. Conventional studies on stratified shear instabilities make Boussinesq approximation. Free-surface, which arises due to large density variations between air and water, cannot be taken into consideration under Boussinesq approximation. Hence the free-surface is usually replaced by a ``rigid-lid'', and therefore has less significant effect on the stability of the fluid below it. In this paper we have performed non-Boussinesq linear stability analyses of a double circulation velocity profile prevalent in two-layered density stratified lakes. 
One of our analyses is performed by considering the presence of wind, while the other one considers quiescent air. Both analyses have shown similar growth rates and stability boundaries. We have compared and contrasted  our non-Boussinesq study with a corresponding Boussinesq one (in which the free-surface is replaced by  a rigid-lid).  The maximum non-Boussinesq growth rate is found to be an order of magnitude greater than the maximum Boussinesq growth rate. Furthermore, the stability curves in these two studies are very different. The non-Boussinesq instability as well as the Boussinseq one can become three dimensional in some sub-ranges of the bulk Richardson number. An analytical study has also been conducted on a simple broken-line profile representing double circulation in order to complement the numerical stability analysis. The analytical growth rates are in  good agreement with the non-Boussinesq numerical growth rates observed in continuous profiles. The physical mechanism behind this non-Boussinesq instability is the resonant interaction between a surface vorticity-gravity wave existing at the free-surface (air-water interface), and an interfacial vorticity-gravity wave existing at the pycnocline (warm  and cold water interface). We expect similar kind of non-Boussinesq instability to occur in the upper layer of oceans.   

\end{abstract}

\section{Introduction}
Waves are ubiquitous in nature and they play a very important role in different geophysical phenomena observed. Waves in nature are present in the form of tsunami waves, capillary-gravity waves, Rossby waves, and so on \cite[]{craik1988wave}.  There is a widespread occurrence of stable density stratification in oceans, estuaries, lakes, terrestrial and planetary atmospheres. A stable density interface supports neutrally propagating interfacial gravity waves. In presence of a background velocity shear, it is possible for a stable density stratified flow (as well as homogeneous flow) to become unstable. The resulting instabilities are known as stratified shear instabilities, and a few well known examples are Kelvin-Helmholtz instability, Holmboe instability, and Taylor-Caulfield instability. Interfacial waves, if present in unstable shear flows, are no longer neutral but can grow/decay at an exponential rate. Instabilities and waves play key roles in many geophysical processes and affect our weather and climate \cite[]{vallis2006atmospheric}. Sir G.\, I.\, Taylor in his seminal work \cite[]{taylor1931effect}  has theoretically demonstrated that in presence of linear shear, two interfacial gravity waves at two different interfaces can interact with each other to give rise to an instability, which is now known as the Taylor-Caulfield instability. \cite{taylor1931effect} is one of the earliest pioneering works that provides a physical understanding of shear instabilities.
 Later on, \cite{holmboe1962behavior} has demonstrated that resonant interaction between two or more neutrally propagating interfacial waves leads to exponentially growing shear instabilities. Researchers like \cite{sakai1989rossby}, \cite{baines1994mechanism}, \cite{caulfield1994multiple}  and \cite{heifetz1999counter} have further developed the theoretical understanding of the physical interpretation of shear instabilities as interacting interfacial waves Doppler-shifted by background shear. Interfacial waves exist at an interface. For example, at a density interface (like that between warm water and cold water), there exists two oppositely propagating interfacial gravity waves. Likewise, at a vorticity interface (jump in background vorticity), there exists a vorticity wave (in a rotating reference frame, this would be a Rossby edge wave). For the instability to occur the waves must be phase-locked, i.e.\ stationary with respect to each other, and must have a relative phase difference such that mutual growth occurs. Under this configuration we have exponential or normal mode type instability. A detailed review of the basics of the wave interaction approach to shear instabilities can be found in \cite{carpenter2011instability}. The physical insight into the mechanisms responsible for instabilities enable us to better understand why instabilities occur and provide us with ways to control the instabilities. Furthermore this physical insight also helps one to predicts when wave interaction can give rise to shear instabilities. In recent years \cite{guha2014wave} have presented a generalized theory of wave interactions in which the wave type (i.e.\ vorticity wave, gravity wave, etc.) is kept arbitrary. They have also provided a necessary and sufficient condition for two interfacial waves with arbitrary initial conditions to phase-lock, resonate and grow exponentially.

\begin{figure}
  \centering{\includegraphics[width=5.25in]{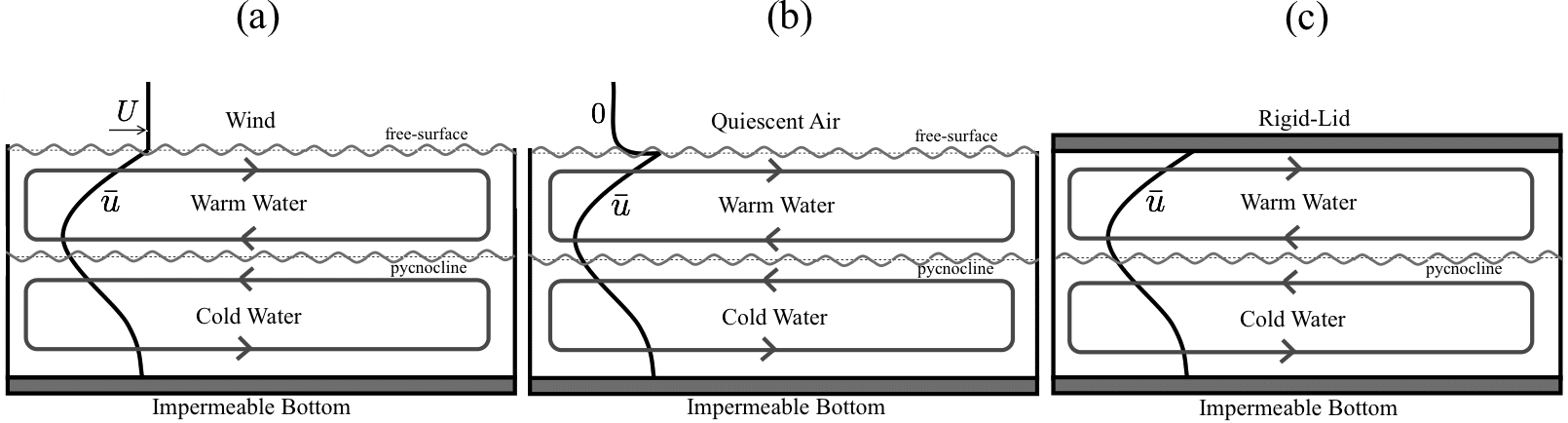}}
  \caption{Each schematic diagram denotes a two-layered density stratified, long and narrow lake undergoing double circulation. (a) Non-Boussinesq case in presence of wind. Here the free-surface (air and lake interface) has been taken into consideration. (b) Non-Boussinesq case in presence of quiescent air. Here also the free-surface has been taken into consideration. (c) Boussinesq case, where the free-surface is replaced by a rigid-lid. }
\label{fig:0}
\end{figure}

Conventionally shear instabilities have been studied in an infinite extent of background fluid. Infinite extent studies do not consider the free-surface of air and water, hence they make a good use of the Boussinesq approximation. Boussinesq approximation neglects the effect of variation of density in the inertial terms, the effect of density variation is considered only in the buoyancy term where the effect is amplified by gravity \cite[]{turner1979buoyancy}. Also the Boussinesq approximation gives accurate results only when the density differences are small as compared to the mean background density. There have been numerous studies on linear stability analyses and direct numerical simulations of stratified shear instabilities, e.g.\ \cite{smyth1988finite}, \cite{smyth1989transition}, \cite{smyth1991instability}, \cite{lawrence1991stability}, \cite{sutherland1992stability}, \cite{alexakis2005holmboe}, \cite{carpenter2007evolution}, \cite{smyth2007mixing}, \cite{carpenter2010identifying}, \cite{guha2013evolution} and \cite{rahmani2014}. These studies have modeled the free-surface as a rigid-lid, and furthermore, they consider very large domain extent as compared to the shear layer thickness (to emulate an infinite domain). 
Relatively few studies have analyzed the effect of domain extent on stability of shear layers. \cite{hazel1972numerical} and \cite{haigh1999symmetric} have shown that the presence of rigid-lid close to the shear layer markedly affects the stability characteristics. In the present study we have not only considered finite extent of the domain but also the effect of free-surface (instead of rigid-lid) on the stability characteristics. Our work is motivated by the fact that most naturally occurring flows have shallow depth and contain a free-surface (interface between air and water).  To the best of our knowledge, effect of free-surface on the stability of the fluid below has not been reported in literature. It is important to note here that a free-surface supports surface gravity waves. As discussed in the previous paragraph, wave interactions lead to shear instabilities. The wave interaction theory of instabilities proposed by \cite{guha2014wave} has shown that the interaction between any two types of interfacial waves can possibly lead to shear instabilities. Thus the interactions between a surface wave and an interfacial wave at the pycnocline (Boussinesq interface between warmer water and colder water) can affect the stability characteristics of the flow. This is especially true for shallow flows shown in  Fig.\ \ref{fig:0}(a) and Fig.\ \ref{fig:0}(b); even moderately long waves at the free-surface and the pycnocline can feel each other's presence and therefore can interact. However, such interactions are not possible if Boussinesq approximation is made. In this case the dynamics of the free-surface cannot be captured, and is therefore replaced by a rigid-lid; see Fig.\ \ref{fig:0}(c). 

Natural flows, e.g.\ lakes, estuaries, oceans, etc.\ are shallow, contain a free-surface, and are density stratified.
Including the effect of free-surface automatically implies that one cannot consider Boussinesq approximation. This is simply because the free-surface (air-water interface) is a non-Boussinesq interface. The traditional approach to modeling surface waves  (specifically long waves like solitary/tsunami waves) is through shallow water approximation \cite[]{vallis2006atmospheric}, in which the surface elevation also  becomes a prognostic variable. Variables like density and horizontal velocity in shallow water approximation are vertically homogeneous (does not vary with depth). Hence shallow water approximation,  although being a very useful technique, is unsuitable for studying  stratified shear instabilities, e.g.\ those occurring in lakes and upper oceans \footnote{\cite{balmforth1999shear} has performed  stability analysis of a shear flow in shallow water, however the shear here is in the cross-stream direction.}.  

There has been studies which considered interaction between surface gravity and interfacial gravity waves in the non-linear and weakly non-linear regimes; see  \cite{ball1964energy}, \cite{jamali2003note}, \cite{jamali2003asymptotic} and \cite{alam2012new}. The instability mechanism occurring here is through wave-triad interactions, for example, interaction between a surface wave and two sub-harmonic interfacial waves, and is essentially a non-linear phenomena. Unlike linear instability, the growth here is algebraic instead of exponential. Moreover, background shear is unimportant for these instabilities, and is therefore absent in the studies mentioned above. In the present study we have shown that a surface wave and an interfacial wave can interact linearly in presence of background shear  (see \S 3 and \S 5) giving rise to exponential growth (i.e.\ linear normal mode instability).

The goal of the current study is to consider shallowness (or finite depth), density stratification and free-surface all together. This would enable us to obtain a more realistic understanding of shear instabilities in natural flows. A linear, non-Boussinesq model  has been developed for this purpose. The non-Boussinseq model captures the effect of large density differences (that between air and water) by considering the effect of density variation in the inertial terms under the purview of linear approximation. 
Similar non-Boussinesq models under inviscid and non-diffusive limit have been studied by \cite{barros2011holmboe} and \cite{heifetz2015stratified}. While \cite{barros2011holmboe} have studied non-Boussinesq Holmboe instability in presence of a large density jump, \cite{heifetz2015stratified} have made use of the non-Boussinesq formulation to arrive at a mechanistic interpretation of non-Boussinesq Taylor-Caulfield instability.

In the present study we have demonstrated the effect of surface wave (at the free-surface) on  interfacial wave (at the pycnocline)  in a long and narrow rectangular lake. Many simplifying assumptions are possible for lakes which are long and narrow. Coriolis effect, which only becomes important when the width of the lake is comparable to the Rossby deformation radius, can be ignored for narrow lakes. If the horizontal dimension is much longer than the depth (i.e.\ shallow flow), the background flow in a large extent of the domain can be assumed to be parallel. This is because the sidewall effects can be neglected away from the boundaries. Previous studies on the effect of wind stresses on lakes have been conducted by \cite{bye1965wind} and \cite{heaps1966wind}. \cite{bye1965wind} has theoretically derived a velocity profile in an unstratified lake, that is in good agreement with laboratory experiments. \cite{heaps1966wind} have analyzed the effect of wind stresses on a two layer stratified lake and have derived a velocity profile for circulation in upper layer. They have presented the effects of surface seiches on the internal seiche at the pycnocline. Our study considers the stability of the base state profiles  in a lake, while the previous studies have been mainly concerned with the setting up of the base state profiles. \cite{hutter2011physics} have presented a theoretical derivation of the base state profile in a two-layered density stratified lake. 
\cite{csanady1975hydrodynamics} in his review article has discussed the hydrodynamics of large-scale motions like Kelvin waves, Poincar\'e waves and the effect of Coriolis parameter on large lakes. In another review article, \cite{imberger1982dynamics} have discussed  different mixing mechanisms and gravitational adjustments that are at work in lakes. Recently \cite{wuest2003small} has payed detailed attention to small scale phenomena  occurring in lakes. \cite{wuest2003small} has also discussed the effect of turbulent mixing, transfer of energy form large-scale motion to turbulence through the bottom boundary layer and the dynamics and interaction of wave affected energetic surface layer. In contrast to the previous works on physical limnology, the current study aims at understanding the linear stability of the base flow, which is set-up by the wind forcing, in two-layered density stratified lakes. Moreover, we emphasize on the role played by the waves on the free-surface  on the stability characteristics. Current work has significance to the limnological community primarily because it paves the way for new types of non-Boussinesq instability mechanisms and subsequent turbulent mixing that can arise inside the lakes. 

The outline of this paper is as follows. In \S 2 the governing equations of the problem  are discussed. A comprehensive derivation of the viscous, diffusive and non-Boussinesq Taylor-Goldstein equation is provided in  Appendix \ref{appA}. A brief account of wind stress induced double circulation in two-layered lakes is presented in \S 3. This section further presents a non-Boussinesq numerical stability of the double circulation velocity profile in the presence of wind as well as quiescent air. 
To contrast our non-Boussinesq studies, we have also performed a Boussinesq stability analysis in \S \ref{sec:4}. Here we have assumed the free-surface to be a rigid-lid. 
Finally in \S 5 we have carried out stability analysis of a simple broken-line profile that captures the essential features of the double circulation profile. 
Final conclusions and summary are presented in \S 6.

\section{Governing Equations and Eigenvalue Solution}\label{sec:2}
\subsection{Governing Equations}
To model a long and narrow lake we use the 2D incompressible Navier-Stokes equations along with the incompressible mass continuity equation.
The  flow is along the $x$-$z$ plane, and $u$ and $w$ are the respective velocity components in the horizontal ($x$) and vertical ($z$) directions. 

The incompressible mass continuity is given by the divergence free velocity condition
\begin{equation}\label{eq:1}
\frac{\partial u}{\partial x}+\frac{\partial w}{\partial z}=0,
\end{equation}
 The $x$ and $z$ momentum equations are respectively given by
\begin{equation}\label{eq:2}
\rho \Big(\frac{\partial u}{\partial t}+u\frac{\partial u}{\partial x}+w\frac{\partial u}{\partial z}\Big)=-\frac{\partial p}{\partial x}+\mu\Big(\frac{\partial^{2}u}{\partial x^{2}}+\frac{\partial^{2}u}{\partial z^{2}}\Big),
\end{equation}
and
\begin{equation}\label{eq:3}
\rho \Big(\frac{\partial w}{\partial t}+u\frac{\partial w}{\partial x}+w\frac{\partial w}{\partial z}\Big)=-\frac{\partial p}{\partial z}-\rho g+ \mu\Big(\frac{\partial^{2}w}{\partial x^{2}}+\frac{\partial^{2}w}{\partial z^{2}}\Big).
\end{equation}
Here $p$ and $g$ respectively denote  pressure and acceleration due to gravity, while $\mu$ and $\rho$ are the dynamic viscosity and the (mass) density of the fluid respectively. Time is given by $t$. 

Advection-diffusion equation for a stratifying agent, e.g.\ temperature or salinity, is given by
\begin{equation}\label{eq:4}
\frac{\partial \theta}{\partial t}+u\frac{\partial \theta}{\partial x}+w\frac{\partial \theta}{\partial z}=\eta\Big(\frac{\partial^{2}\theta}{\partial x^{2}}+\frac{\partial^{2}\theta}{\partial z^{2}}\Big).
\end{equation}
Here $\theta$ represents the stratifying agent while $\eta$ is its molecular diffusivity. We use a linear equation of state to relate  $\theta$ with  $\rho$:
\begin{equation}\label{eq:5}
\rho =\rho_{0}[1-\gamma(\theta-\theta_{0})].
\end{equation}
The coefficient $\gamma$ linearly relates changes in stratifying agent to that of density. Using the above state equation and the advection-diffusion equation, we arrive at an advection-diffusion equation for the density:
\begin{equation}\label{eq:6}
\frac{\partial \rho}{\partial t}+u\frac{\partial \rho}{\partial x}+w\frac{\partial \rho}{\partial z}=\kappa\Big(\frac{\partial^{2}\rho}{\partial x^{2}}+\frac{\partial^{2}\rho}{\partial z^{2}}\Big),
\end{equation}
where $\kappa$ is the diffusivity of density.

 We assume a base state that varies only along $z$, and is given as follows: $u=\bar{u}(z)$, $w=0$, $p=\bar{p}(z)$ and $\rho=\bar{\rho}(z)$. The base state follows  hydrostatic balance $d\bar{p}/dz=-\bar{\rho}g$. Perturbations are then added to the base state: $u=\bar{u}(z)+\tilde{u}$, $w=\tilde{w}$, $p=\bar{p}(z)+\tilde{p}$, and $\rho=\bar{\rho}(z)+\tilde{\rho}$, where $\tilde{f}$ denotes the perturbation quantities. Assuming perturbations are infinitesimal,   (\ref{eq:1})-(\ref{eq:6}) are then linearized about the base state. Furthermore, the perturbations are  assumed to be of the temporal normal mode form: $\tilde{f}=\hat{f}(z)\ee^{\ii\alpha(x-ct)}$. Here  $\alpha$ and $c$ are respectively the real wavenumber and the complex phase speed ($c = c_{r}+\ii c_{i}$), and $f$ could represent  $u,\,w,\,\psi$,\, $p$ or $\rho$. Finally we obtain a system of two equations in terms $\hat{w}$ and $\hat{\rho}$:
\begin{equation}\label{eq:7} 
\bar{\rho}^{\prime}[-(\bar{u}-c)\hat{w}^{\prime}+\bar{u}^{\prime}\hat{w}]+\bar{\rho}[-(\bar{u}-c)\hat{w}^{\prime\prime}+\alpha^{2}(\bar{u}-c)\hat{w}+ \bar{u}^{\prime\prime}\hat{w}]=\ii\alpha\hat{\rho}g+\frac{\ii}{\alpha}\mu[\hat{w}^{\prime\prime\prime\prime}-2\alpha^{2}\hat{w}^{\prime\prime}+\alpha^{4}\hat{w}],
\end{equation}

\begin{equation}\label{eq:8}
\ii \alpha(\bar{u}-c) \hat{\rho}+\hat{w}\bar{\rho}^{\prime}=\kappa(\hat{\rho}^{\prime\prime}-\alpha^{2}\hat{\rho}).
\end{equation}
In (\ref{eq:7})-(\ref{eq:8}), the symbol $^{\prime}$ denotes an ordinary derivative with respect to $z$,  $\hat{\rho}$ and $\hat{w}$ are respectively the eigenfunctions of the perturbation density and vertical velocity, while $\bar{u}$ and $\bar{\rho}$ are the base state horizontal velocity and  density respectively. The above equation set is basically the viscous, diffusive and non-Boussinesq version of the celebrated Taylor-Goldstein equation.
The original Taylor-Goldstein equation, derived independently by \cite{taylor1931effect} and \cite{goldstein1931stability},  is applicable in the Boussinesq, inviscid and non-diffusive limit of (\ref{eq:8}). Hereafter, (\ref{eq:7}) and (\ref{eq:8}) will be referred to as the viscous diffusive non-Boussinesq Taylor-Goldstein equation. A detailed derivation of this equation has been provided in Appendix \ref{appA}. The inviscid and non-diffusive form of  (\ref{eq:7})-(\ref{eq:8}) can be found in \cite{barros2011holmboe}, \cite{barros2014elementary} and \cite{carp2016}.
\subsection{Numerical Solution of Eigenvalue Problem}
Equations (\ref{eq:7}) and (\ref{eq:8}) can be rewritten in the from of a generalized eigenvalue problem of the form of $AX=\lambda BX$ :

\begin{equation}\label{eq:9}
\begin{bmatrix} 
A_{11} & A_{12} \\
A_{21} & A_{22} 
\end{bmatrix}
\begin{bmatrix}
X_{1} \\
X_{2}
\end{bmatrix}
=c
\begin{bmatrix}
B_{11} & B_{12} \\
B_{21} & B_{22} 
\end{bmatrix}
\begin{bmatrix}
X_{1} \\
X_{2}
\end{bmatrix}.
\end{equation}

The terms of the matrices are given by 
\begin{equation*}
A_{11}=\bar{\rho}^{\prime}[-\bar{u}D + \bar{u}^{\prime}] + \bar{\rho}[-\bar{u}D^{2} + \alpha^{2}\bar{u} + \bar{u}^{\prime\prime}]- \frac{\ii}{\alpha}\mu[D^{4}-2\alpha^{2}D^{2} + \alpha^{4}],
\end{equation*}
\begin{equation*}
A_{12}=-\ii\alpha g,\,\,\,\,A_{21}=\bar{\rho}^{\prime},\,\,\,\,A_{22}=\ii\alpha \bar{u} - \kappa[D^{2}-\alpha^{2}],\,\,\,\,X_{1}=\hat{w},\,\,\,\,X_{2}=\hat{\rho},\,\,\,
\end{equation*}
\begin{equation*}
B_{11}=-\bar{\rho}^{\prime}D+\bar{\rho}[-D^{2}+\alpha^{2}],\,\,\,\,B_{12}=0,\,\,\,\,B_{21}=0,\,\,\,\,B_{22}=\ii\alpha.\,\,\,\,\,\,\,\,\,\,\,\,\,\,\,\,\,\,\,\,\,\,\,\,\,\,\,\,\,
\end{equation*}

Equation (\ref{eq:9}) forms a matrix eigenvalue problem for the eigenvalue $c$ and the eigenvectors $\hat{w}$ and $\hat{\rho}$. Here $D$, $D^{2}$ and $D^{4}$ are the first, second and fourth derivative matrices. For calculating $D$, $D^{2}$ and $D^{4}$ we  use the fourth order central difference scheme. The boundary points are discretized using second order one-sided finite difference scheme.

The boundary conditions used for $\hat{w}$ are  impenetrability and free-slip. The impenetrable boundary condition arises due to the continuum hypothesis and is given by
\begin{equation}\label{eq:10}
\hat{w}=0.
\end{equation}
The free-slip boundary condition is given by
\begin{equation}\label{eq:11}
\frac{d^{2}\hat{w}}{dz^{2}}=0.
\end{equation}
The boundary condition used for $\hat{\rho}$ is the insulating boundary condition given by
\begin{equation}\label{eq:12}
\frac{d\hat{\rho}}{dz}=0.
\end{equation}
These boundary conditions are implemented using second order finite difference schemes. Equation (\ref{eq:9}) along with the boundary conditions (\ref{eq:10})-(\ref{eq:12}) form the matrix eigenvalue problem, which we solve using the inbuilt functions in MATLAB. The solution procedure used is similar to \cite{smyth2011narrowband}. We have validated our code for the Bousinesq shear layer problem described in \cite{smyth1988finite}.
Using the matrix eigenvalue solver we have analyzed instabilities in the  velocity and density profiles given in Fig.\ \ref{fig:1} and Fig.\ \ref{fig:6}. In \S \ref{sec:3} and \S \ref{sec:4} we have presented the results of the stability analyses.

\section{Stability of Non-Boussinesq Continuous Profiles}\label{sec:3}
\subsection{Continuous Profiles for Long and Narrow Lakes}
Analytical solutions for the velocity profile in two-layer density stratified rectangular lakes undergoing a steady  wind forcing are possible, the detailed solution procedure is outlined in \cite{hutter2011physics}. Here we will present the results along with some underlying assumptions.
A rectangular basin is considered for simplicity.  The lake is considered to be narrow in order to ignore the effect of Coriolis force. Moreover, the length of the lake is considerably longer than its depth, ensuring that the flow at the center of the lake is unaffected by the sidewalls. The flow is assumed to be two dimensional, incompressible, and  stratified in two layers. The properties like  density and dynamic viscosity are assumed to be constant in each layer. The wind stresses vary along the length of the lake and are assumed to be time independent. Under the above mentioned simplifying assumptions it is possible to obtain a closed-form velocity profile in a two-layered density stratified lake. The non-dimensional base state horizontal velocity profile $\bar{u}$ in terms of the non-dimensional vertical coordinate $z$ is given below:

\begin{equation}\label{eq:13}
\bar{u}(z) = \left\{
        \begin{array}{cc}
            \frac{15}{4}\Big(\frac{2h-z}{h}\Big)^{2} -6\Big(\frac{2h-z}{h}\Big) + \frac{7}{4} & \quad 2h \leq z \leq h, \\ \\
            -\frac{3}{4}\Big(\frac{h-z}{h}\Big)^{2} + \frac{3}{2}\Big(\frac{h-z}{h}\Big)-\frac{1}{2}  & \quad h \leq z \leq 0.
        \end{array}
    \right.
\end{equation}
Here $h$ is the lake half-width. The boundary conditions used to obtain this velocity profile are no-slip at the pycnocline and free-slip  between the heavier lower layer and the bottom topography.

The horizontal velocity profile given by (\ref{eq:13}) is inside the lake domain. However, if the effect of the free-surface is to be considered, the horizontal velocity of the air above has to be taken into account.  Hence we have to extend the profile in (\ref{eq:13}) into the air region, which has been done as follows:

\begin{equation}\label{eq:14}
\bar{u}(z) = \left\{
        \begin{array}{cc}
        \frac{7}{4} + (0.05)\tanh(24(z-2h)) & \quad 2.4h \leq z \leq 2h, \\ \\
            \frac{15}{4}\Big(\frac{2h-z}{h}\Big)^{2} -6\Big(\frac{2h-z}{h}\Big) + \frac{7}{4} & \quad 2h \leq z \leq h, \\ \\
            -\frac{3}{4}\Big(\frac{h-z}{h}\Big)^{2} + \frac{3}{2}\Big(\frac{h-z}{h}\Big)-\frac{1}{2}  & \quad h \leq z \leq 0.
        \end{array}
    \right.
\end{equation}
The extension of the velocity profile into the air region has been done such that (i) the profile remains constant with $z$ away from the free-surface, and (ii) the profile is continuously differentiable at the free-surface. The first point eliminates the possibility of the 
development of a critical layer inside the air region (and therefore wind-driven instability of \cite{miles1957generation} cannot occur). The second point eliminates the possibility of Gibb's phenomena in numerical stability analysis.

\begin{figure}
  \centering{\includegraphics[width=4in]{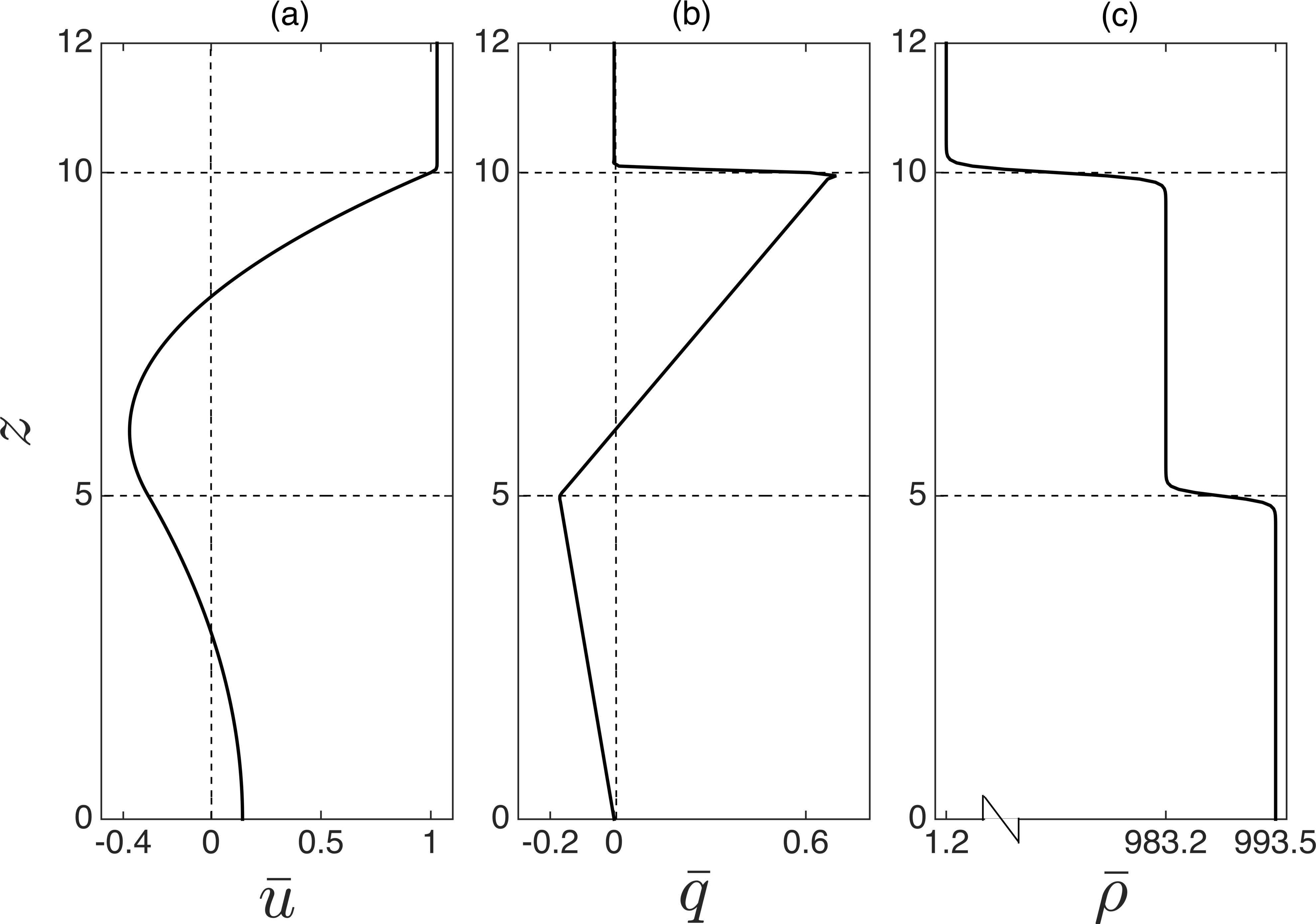}}
  \caption{Base state profiles for then non-Boussinesq case in presence of wind: (a) Base state velocity profile. (b) Base state vorticity profile. (c) Base state density profile for $At=0.01$.}
\label{fig:1}
\end{figure}

The plot for the $u$ profile given in (\ref{eq:14}) is shown in Fig.\ \ref{fig:1}(a). The base state vorticity, $\bar{q}=d\bar{u}/dz$, is given in  Fig.\ \ref{fig:1}(b). As mentioned previously, we are considering two-layered density stratified lakes, meaning each layer of water has a constant density. Moreover density of air region is also constant. Since jump in density will result in Gibb's phenomena in numerical stability analysis, we replace the two-layered lake density profile by a density profile that is continuously differentiable at both the free-surface  as well as the pycnocline. Moreover the profile chosen is such that density is constant in each layer away from the interfaces:   
\begin{equation}\label{eq:15}
\bar{\rho}(z) = \left\{
        \begin{array}{cc}
            \rho_{1} - \frac{\Delta\rho_{1}}{2}\tanh(10(z-2h))  & \quad 2.4h \leq z \leq 1.5h, \\ \\
            \rho_{2} - \frac{\Delta\rho_{2}}{2}\tanh(10(z-h))  & \quad 1.5h \leq z \leq 0.
        \end{array}
    \right.
\end{equation}
This density profile is shown in Fig.\ \ref{fig:1}(c). We can define two Atwood numbers for the system that we have considered. At the free-surface, i.e.\ the interface of air and warm water we have an Atwood number defined by 
\begin{equation}
At_{S}=\frac{2(\rho_{water}-\rho_{air})}{\rho_{water} + \rho_{air}}.
\end{equation}
Since the density of air is orders of magnitude less than the density of water we always have  $At_{S} \rightarrow1$. Atwood number corresponding to the pycnocline, i.e.\ the interface of warm water and cold water is given by 
\begin{equation*}
At=\frac{\Delta\rho_{2}}{\rho_{2}}.
\end{equation*}
Only this Atwood number would vary and is thus a significant parameter of the problem.
\begin{table}
  \begin{center}
\def~{\hphantom{0}}
  \begin{tabular}{lccccc}
      $At$    &$\rho_{1}$ & $\Delta\rho_{1}$ & $\rho_{2}$ & $\Delta\rho_{2}$ & $h$    \\[3pt]
       \, & ($\frac{Kg}{m^{3}}$) & ($\frac{Kg}{m^{3}}$) & ($\frac{Kg}{m^{3}}$) & ($\frac{Kg}{m^{3}}$) &(m) \\~\\
      0.01 & 492.2 & 982 & 988.35 & 10.3  & 5\\
     0.003 & 498.45 & 994.5 & 997.42 & 3.44 & 5\\
       
  \end{tabular}
  \caption{Representative values of Atwood number, mean densities and density jumps at the free-surface and the pycnocline, and  lake half-width.}
  \label{tab:first}
  \end{center}
\end{table}
Table \ref{tab:first} provides the values of Atwood numbers, the mean densities and density jumps at the free-surface and the pycnocline, as well as the half-width $h$ of the lake  used in this paper.  Fig.\ \ref{fig:1}(b) indicates that a vorticity wave (which exists at the jump of base state vorticity) can be supported at the free-surface, while  Fig.\ \ref{fig:1}(c) indicates that gravity waves can be supported both at the free-surface as well as the pycnocline. Therefore,  surface vorticity-gravity waves can be present at the free-surface, while interfacial gravity waves can be present at the pycnocline. 

\subsection{Stability of Continuous Profiles in Presence of Wind}
\label{sub:wind}
In this sub-section we present the results for the case depicted in Fig. \ref{fig:0}(a). For the results discussed in this sub-section, we have non-dimensionalized the wavenumber, $\alpha$, the phase speed, $c$, and the growth rate $\Im\{\alpha c\}$. The wavenumber is non-dimensionalized by $h$ while the phase speed is non-dimensionalized by the velocity scale, $U$. These quantities are also used to non-dimensionalize the growth rates.
\begin{figure}
  \centering{\includegraphics[width=5in]{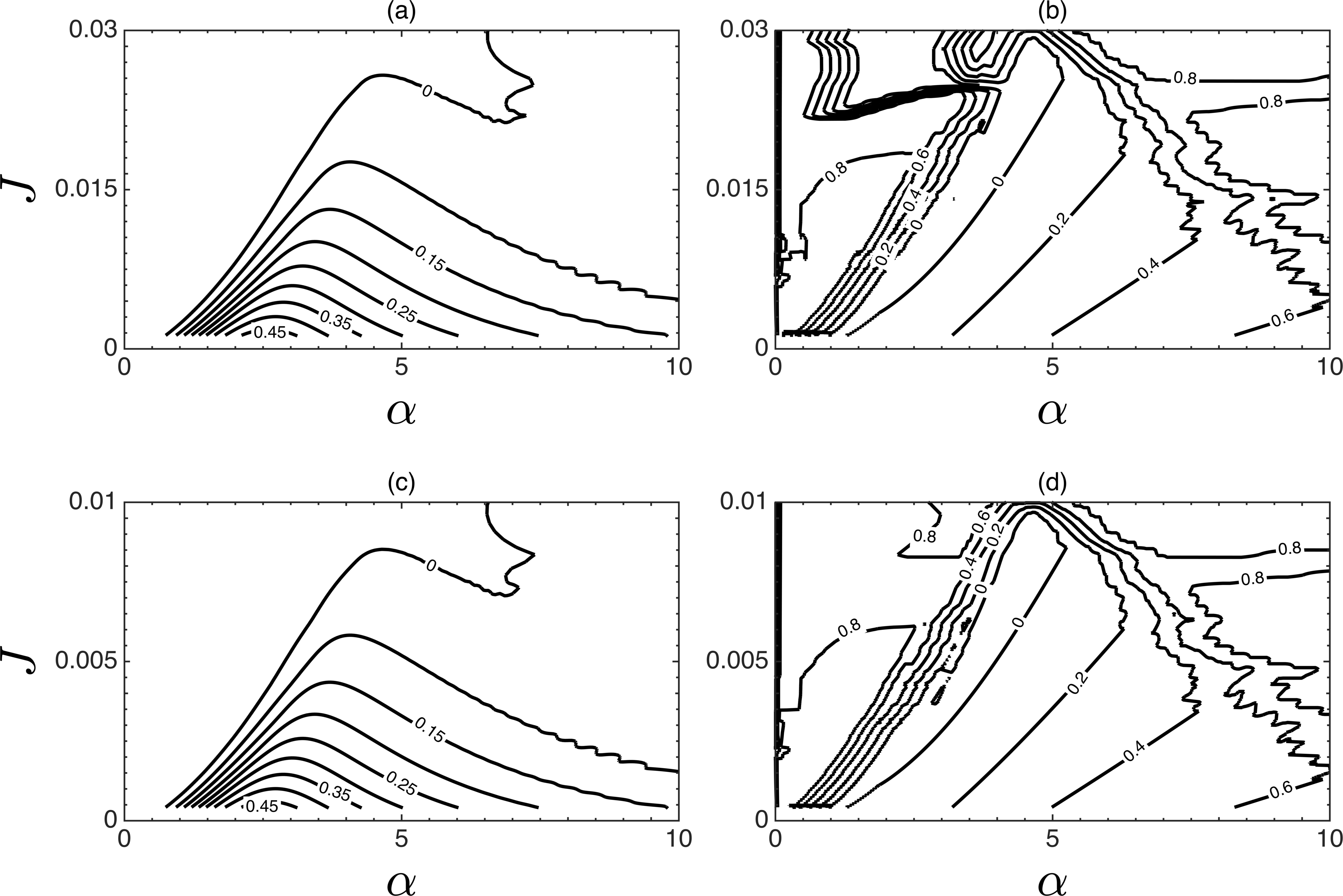}}
  \caption{Non-Boussinesq growth rates and phase speeds for the wind flow case: (a) Growth rates for $At=0.01$, (b) phase speeds for $At=0.01$, (c) growth rates for $At=0.003$, and (d) phase speeds for $At=0.003$. The stability boundaries for the two Atwood numbers are a scaled version of each other.}
\label{fig:2}
\end{figure}

Figure \ref{fig:2} depicts the growth rates that are observed when the non-Boussinesq continuous profile (shown in Fig. \ref{fig:1}) is numerically tested for instabilities. We have performed the numerical stability analyses for two Atwood numbers given in Table \ref{tab:first}. For both Atwood numbers a maximum growth rate of $0.4821$ occurs at $\alpha = 2.509$ and $J = 0$. As can be seen from Fig.\ \ref{fig:2}(a) and Fig. \ref{fig:2}(c), the two Atwood numbers have the same maximum growth rate, implying that maximum growth rate is independent of $At$. Stability boundary for $At=0.003$ is a scaled down version of the stability boundary for $At=0.01$. The scaling is due to the Atwood number dependence of the bulk Richardson number, $J$, given by
\begin{equation}\label{eq:J}
J=At\frac{gh}{U^{2}}.
\end{equation}
As intuitively expected, keeping $\alpha$ constant and increasing $J$ causes stabilization of the flow. 
The phase speed for the two different Atwood numbers is same in the region of instability as can be seen from Fig.\ \ref{fig:2}(b) and Fig.\ \ref{fig:2}(d), but slight differences are observed in the region outside the stability boundaries.


\begin{figure}
  \centering{\includegraphics[width=5in]{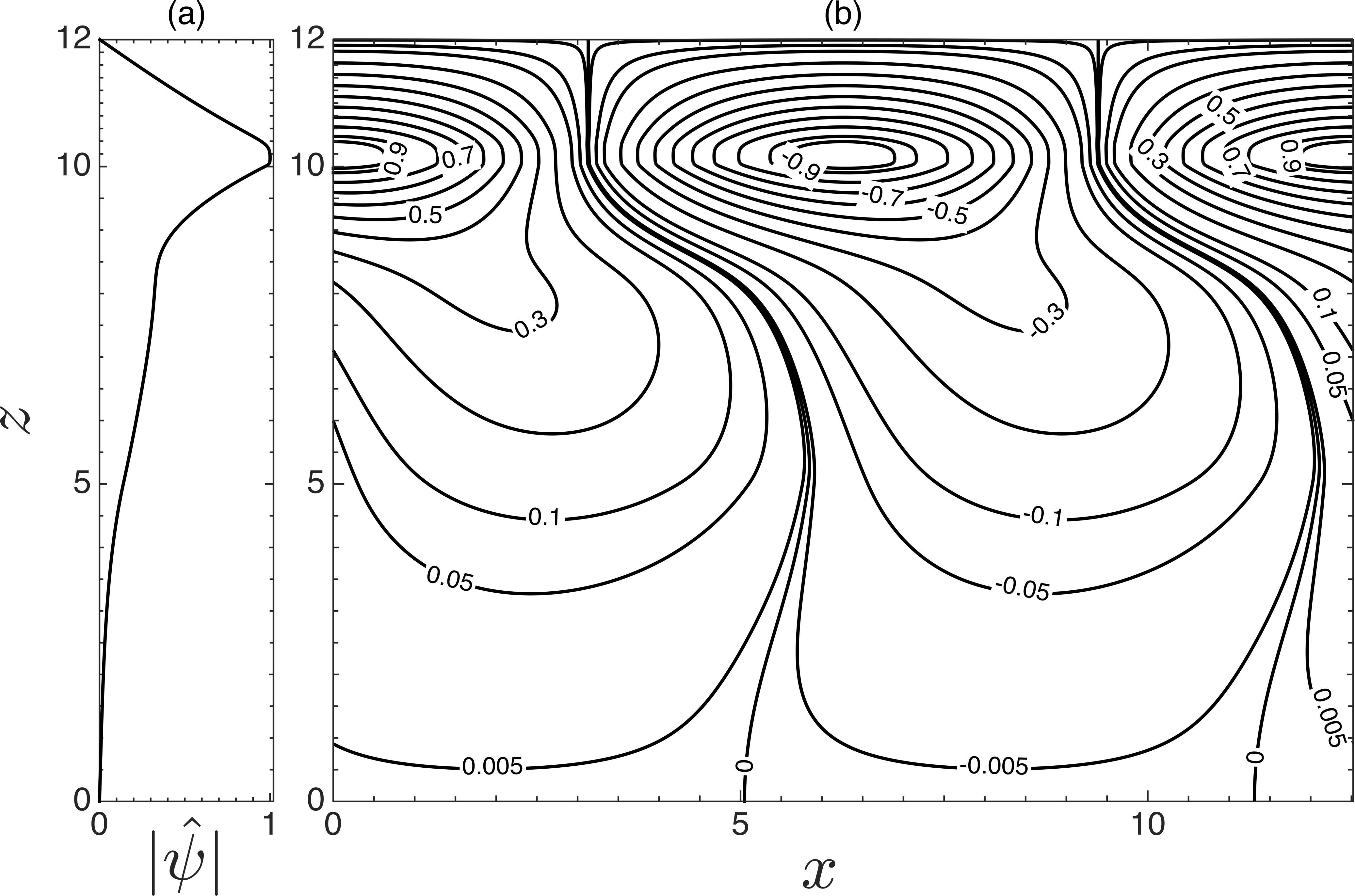}}
  \caption{ Plots for the non-Boussinesq perturbation stream function for wind flow case: (a) Norm of  perturbation stream function eigenfunction versus $z$, and (b) contours of perturbation stream function $\tilde{\psi}$.}
\label{fig:3}
\end{figure}
Eigenfunction of the perturbation stream function, $\hat{\psi}$, is obtained from the vertical velocity eigenfunction, $\hat{w}$, through the relation $\hat{\psi}=-\ii\hat{w}/k$. For reporting we have normalized $\hat{\psi}$ with its maximum value. The quantity $\hat{\psi}$ peaks at the free-surface, which can be seen in Fig. \ref{fig:3}(a). This figure has been drawn corresponding to the  maximum growth rate ($\alpha=2.509$, $J=0$) case. 
The contours of perturbation stream function, $\tilde{\psi}=\Re \{\hat{\psi} \ee^{\ii \alpha x} \}$  for one wavelength of the disturbance is depicted in Fig. \ref{fig:3}(b). 
Tangent at each point on the stream function contour gives the direction of the perturbation velocity field. 

\begin{figure}
  \centering{\includegraphics[width=5in]{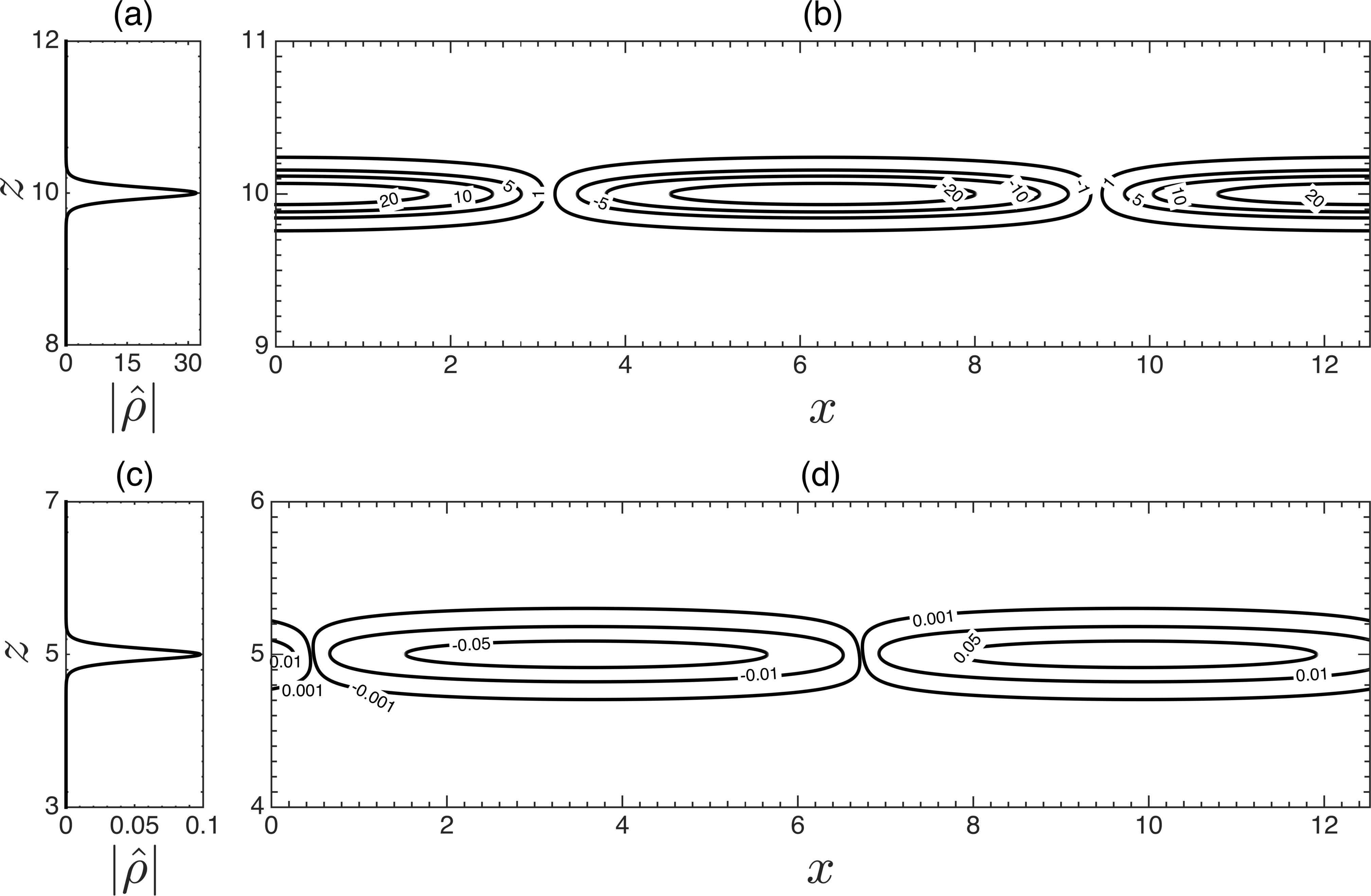}}
  \caption{Plots of non-Boussinesq  density perturbation for wind flow case. The perturbation density eigenfunctions are normalized by the maximum value of corresponding perturbation stream function eigenfunction. (a) Norm of perturbation density eigenfunction versus $z$ near the free surface. (b) Contours of perturbation density, $\tilde{\rho}$, near the free surface. (c) Norm of perturbation density eigenfunction versus $z$ near the pycnocline. (d) Contours of perturbation density, $\tilde{\rho}$, near the pycnocline.}
\label{fig:4}
\end{figure}
 Perturbation density, $\tilde{\rho}$, and the corresponding eigenfunction $\hat{\rho}$, are plotted in Fig.\, \ref{fig:4}. The quantity $\hat{\rho}$ is normalized by the maximum value of the corresponding $\hat{\psi}$.  We find that $\hat{\rho}$ peaks at the free-surface and the pycnocline, the former being few orders of magnitude greater than the latter, as can be observed by comparing Fig.\ \ref{fig:4}(a) with Fig. \ref{fig:4}(c).  This is simply because the density difference at the free-surface is much larger than that at  the pycnocline.  Density contours around the free-surface is depicted in Fig.\ \ref{fig:4}(b), while those around the pycnocline is shown in Fig.\ \ref{fig:4}(d). The former indicates the presence of surface gravity waves, while the latter indicates interfacial gravity waves. Surface gravity wave and the interfacial gravity waves are not far away, and can interact with each other through the perturbed velocity field. As can be inferred from the perturbation streamfunction plot in Fig.\ \ref{fig:3}(b), the perturbed velocity field is non-zero in the region between the two waves.


\subsection{Relationship Between 2D Instabilities and 3D Instabilities}\label{sec:SP3D}
\begin{figure}
  \centering{\includegraphics[width=4.5in]{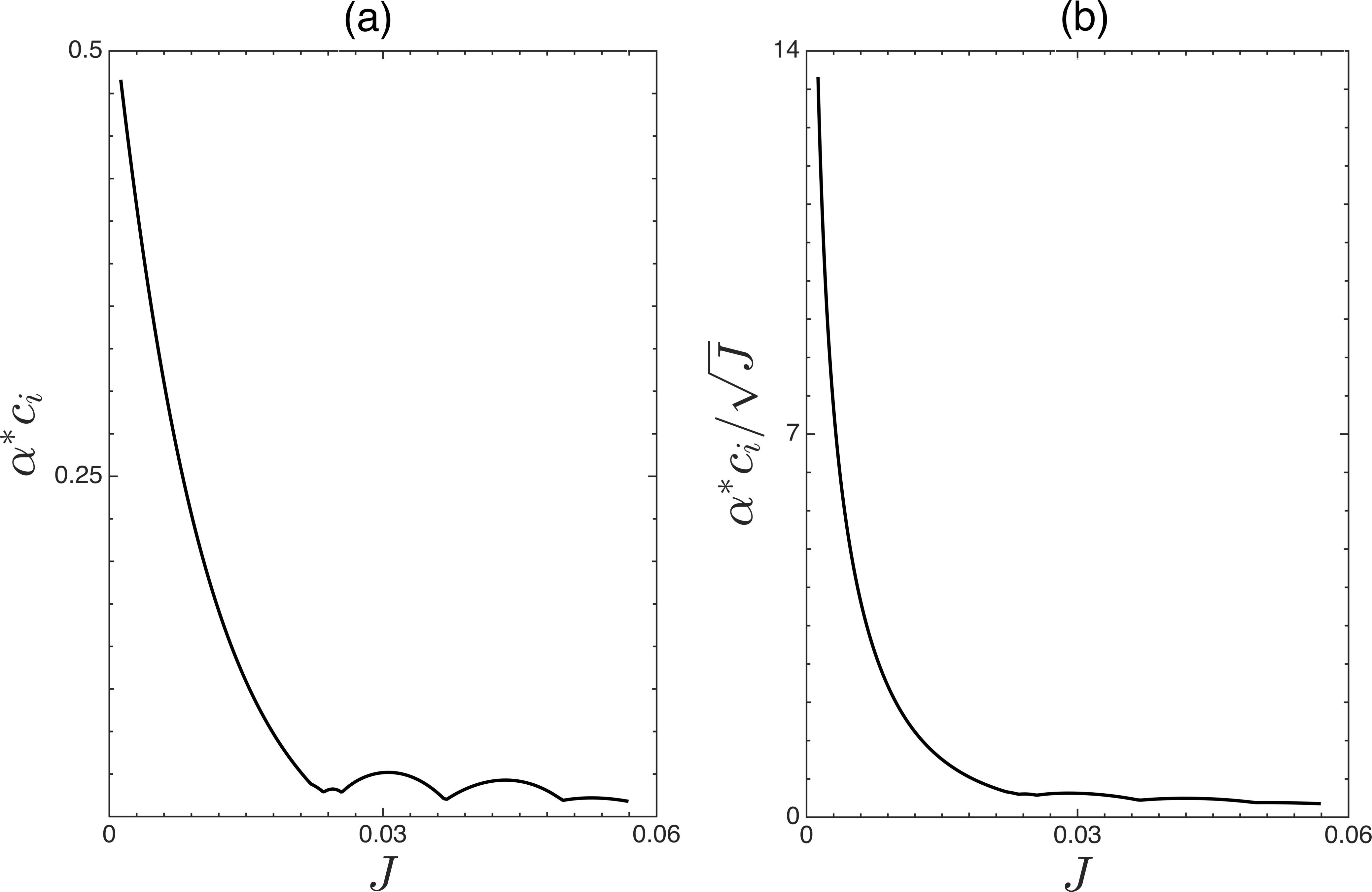}}
  \caption{(a) Variation of maximum growth rate, $\alpha^{*} c_{i}$ (evaluated at each $J$), versus $J$. (b) Plot of $\alpha^{*}c_{i}/\sqrt{J}$ versus $J$. }
\label{fig:5}
\end{figure}

We have discussed two dimensional instabilities until now, but in this sub-section we will elaborate on the relationship between two dimensional and three dimensional inabilities. We follow a procedure similar to \cite{smyth1988finite}, \cite{smyth1990three} and \cite{haigh1995}. To this effect we present a non-Boussinesq Taylor-Goldstein equation that has been obtained after introduction of three dimensional perturbations. A general form of three dimensional perturbation is of a progressive wave propagating at an angle $\phi$ \, to the $x$-axis and having an amplitude varying in $z$ direction. The three dimensional normal mode perturbations take the form $\tilde{f}(x,y,z,t)=\hat{f}(z)\ee^{\ii \alpha(x\cos\phi+y\sin\phi-ct)}$ \cite[]{white2006viscous}. On using this form of disturbances we arrive at a non-Boussinesq Taylor-Goldstein equation. For simplicity we present the inviscid, non-diffusive case:
\begin{equation}\label{eq:s1}
\bar{\rho}^{\prime}\Big[(\bar{u}-\frac{c}{\cos\phi})\hat{w}^{\prime}- \bar{u}^{\prime}\hat{w} - \frac{g}{\cos^{2}\phi(\bar{u}-\frac{c}{\cos\phi})}\hat{w} \Big]  +  \bar{\rho}\Big[(\bar{u}-\frac{c}{\cos\phi})(\hat{w}^{\prime}-\alpha^{2}\hat{w}) - \bar{u}^{\prime\prime}\hat{w} \Big]=0.
\end{equation}
For a detailed derivation of  non-Boussinesq Taylor-Goldstein equation with  three dimensional perturbations, the reader is directed towards Appendix \ref{appB}. A similar equation for Boussinesq case is obtained by \cite{haigh1995}.
Equation (\ref{eq:s1}) is nothing but the non-Boussinesq Taylor-Goldstein equation for two dimensional perturbation given by (\ref{eq:19}), with a scaled gravity term given by $g/\cos^{2}\phi$ and a complex phase speed given by $c/\cos\phi$. Thus, evaluating the growth rate for the three dimensional problem is equivalent to evaluating the growth rate for a related two dimensional problem and scaling it by $\cos\phi$:
\begin{equation}\label{eq:s2}
\alpha c_{i}(\bar{u},\phi,\alpha,g)=\cos\phi.\alpha c_{i}(\bar{u},0,\alpha,g/\cos^{2}\phi).
\end{equation}
Hence we may infer that the most unstable mode in a density stratified parallel shear flow will become three dimensional if the growth rate of the most unstable two dimensional mode increases sufficiently rapidly on increasing $g$. Conversely it can be deduced that the dominant unstable mode will be two dimensional if the growth rate of the fastest growing mode decreases on increasing  $g$. In Fig.\ \ref{fig:5} we report the results with non-dimensional variables, hence $g$ gets non-dimensionalized to $J$. Therefore the ``apparent'' condition for disturbances to remain two dimensional is
\begin{equation}\label{eq:s2.5}
\frac{d(\alpha^{*}c_{i})}{dJ}<0.
\end{equation}
Here $\alpha^{*}$ is the non-dimensional wave number for which maximum growth rate occurs for a given value of $J$.
For the case of non-Boussinesq instability in presence of wind flow we can see that $d(\alpha^{*}c_{i})/dJ$ does not decrease monotonically; see Fig.\ \ref{fig:5}(a).  Care has to be taken because $d(\alpha^{*}c_{i})/dJ>0$ does not necessarily imply three dimensionality of perturbations. This is because of the factor $\cos\phi$ multiplying the growth rate. In the earlier sections we have assumed that the perturbations are initially two dimensional. This assumption is well founded if the fastest growing two dimensional mode is more unstable than the fastest growing three dimensional mode; i.e.
\begin{equation}\label{eq:s3}
\cos\phi.\alpha^{*}c_{i}(\bar{u},0,\alpha,J/\cos^{2}\phi)\leq\alpha^{*}c_{i}(\bar{u},0,\alpha,J).
\end{equation}
 For $J>0$ we get
\begin{equation}\label{eq:s4}
\frac{\cos\phi}{\sqrt{J}}.\alpha^{*}c_{i}(\bar{u},0,\alpha,J/\cos^{2}\phi)\leq\frac{1}{\sqrt{J}}\alpha^{*}c_{i}(\bar{u},0,\alpha,J).
\end{equation}
In order to corroborate the two dimensional assumption, we must show that $\alpha^{*}c_{i}/\sqrt{J}$ decreases with increasing $J$ along the curve of maximum growth rate for a given value of $J$.
In Fig.\ \ref{fig:5}(b) we see that $\alpha^{*}c_{i}/\sqrt{J}$ as a function of $J$ is not a strictly monotonically decreasing function.  Hence we can conclude that the instability is two dimensional in most of the ranges of $J$, but there are sub-ranges where the instability might become three dimensional. 

\subsection{Stability of Continuous Profiles in Presence of Still Air}
\begin{figure}
  \centering{\includegraphics[width=4in]{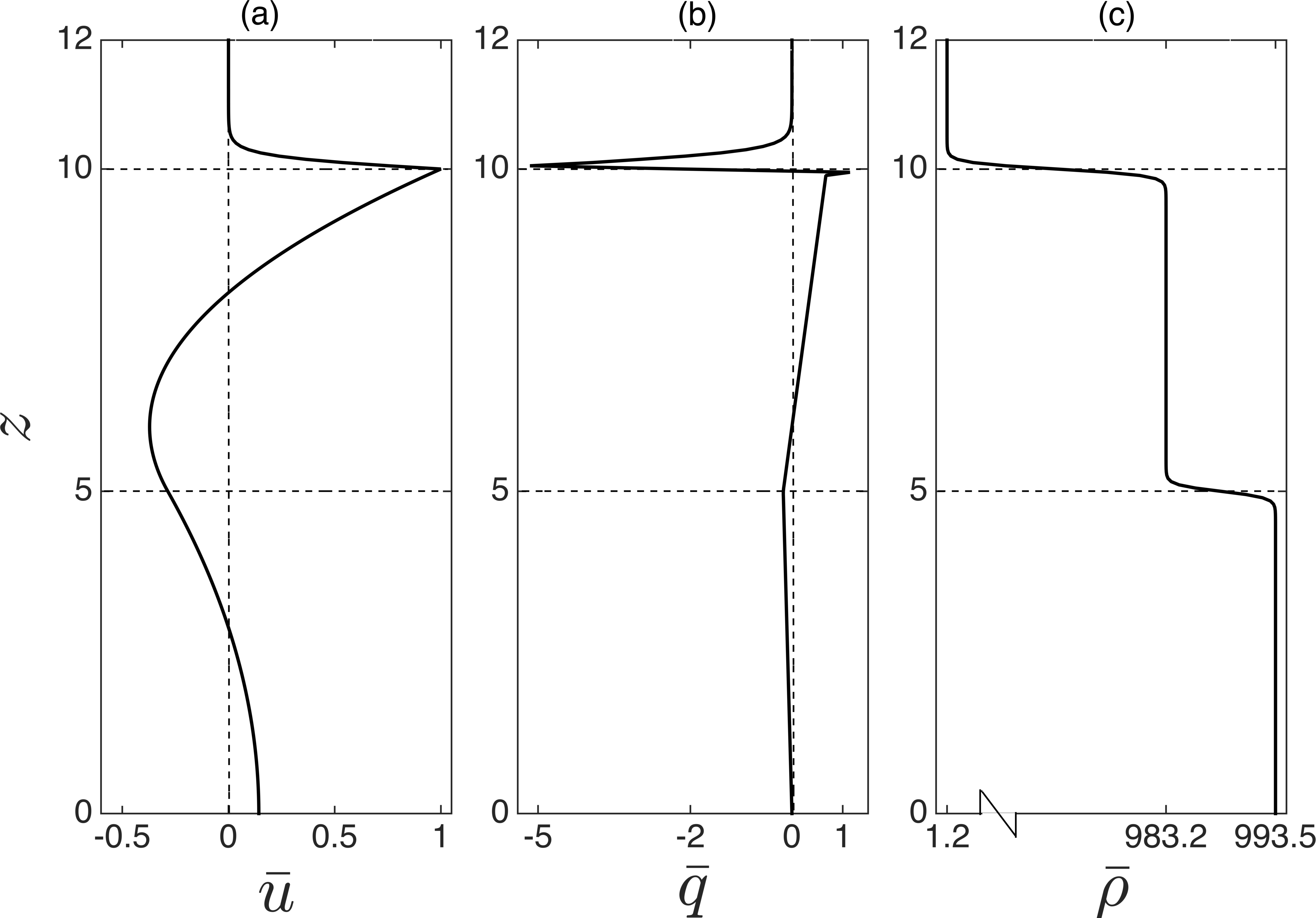}}
  \caption{Base state profiles for the non-Boussinesq case in presence of still air. (a) Base state velocity profile. (b) Base state vorticity profile. (c) Base state density profile for $At=0.01$.}
\label{fig:6}
\end{figure}

In nature, a wind forcing event is responsible for the setting up of the double circulation profile in two-layered density stratified lakes. Previously we have presented  detailed analyses of the instabilities that can occur if the wind forcing event persists after the setting up of the double circulation. The resulting instability arises due to the mutual interaction  between the surface vorticity gravity wave and the interfacial  gravity wave.  Therefore, the free-surface  plays a major role in the occurrence of the instability. In this sub-section a detailed analysis is presented for the case when, after setting up the double circulation profile, the air above the lake reaches a quiescent state. The schematic of this case has been shown in  Fig.\ \ref{fig:0}(b). Previously it has been demonstrated that Atwood number has no effect on the growth rates, Atwood number only scales the bulk Richardson number. Hence in the current sub-section only results corresponding to  $At=0.01$ are presented. The base state velocity profile for the case of quiescent air is given by:
\begin{equation}\label{eq:sa1}
\bar{u}(z) = \left\{
        \begin{array}{cc}
        \frac{7}{4} - \frac{7}{4}\tanh(5(z-2h)) & \quad 2.4h \leq z \leq 2h, \\ \\
            \frac{15}{4}\Big(\frac{2h-z}{h}\Big)^{2} -6\Big(\frac{2h-z}{h}\Big) + \frac{7}{4} & \quad 2h \leq z \leq h, \\ \\
            -\frac{3}{4}\Big(\frac{h-z}{h}\Big)^{2} + \frac{3}{2}\Big(\frac{h-z}{h}\Big)-\frac{1}{2}  & \quad h \leq z \leq 0.
        \end{array}
    \right.
\end{equation}
The density profile for $At=0.01$ is given by:
\begin{equation}\label{eq:sa2}
\bar{\rho}(z) = \left\{
        \begin{array}{cc}
            492.2 - 491\tanh(10(z-2h))  & \quad 2.4h \leq z \leq 1.5h, \\ \\
            988.35 - 5.15\tanh(10(z-h))  & \quad 1.5h \leq z \leq 0.
        \end{array}
    \right.
\end{equation}
In this sub-section we have non-dimensionalized  wavenumber, phase speed and growth rate using the same scales given in \S \ref{sub:wind}. Figure \ref{fig:6} presents the base state profiles for the case in which the air above the lake is stationary. As discussed earlier, the extension of velocity profile into the air region is done such that the profile is continuously differentiable at the free-surface and is constant (equal to zero) away from the free-surface. Similar care has also been taken for the density profile.  

\begin{figure}
  \centering{\includegraphics[width=5in]{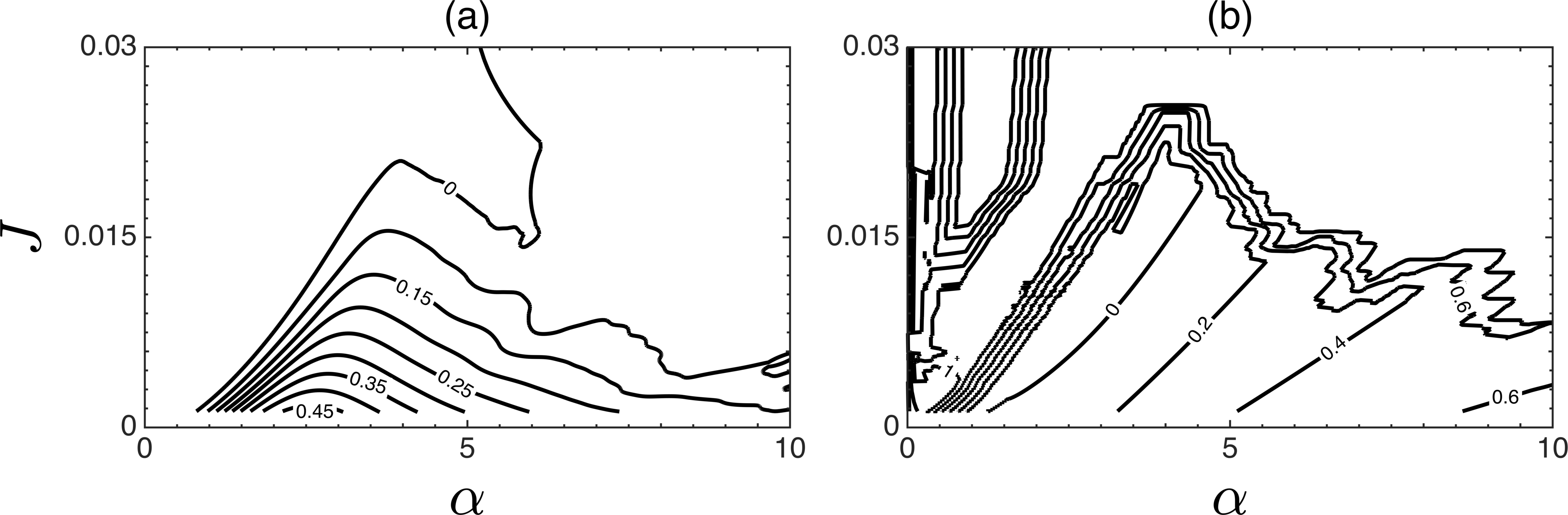}}
  \caption{Non-Boussinesq growth rates and phase speeds in the presence of still air. (a) Growth rates for $At=0.01$. (b) Phase speeds for $At=0.01$. }
\label{fig:7}
\end{figure}

The continuous profiles (Fig. \ref{fig:6}) are numerically tested for instabilities. Figure \ref{fig:7} provides the growth rates and phase speeds that are observed for the case of quiescent air above the lake.  A maximum growth rate of $0.4766$ is observed for $\alpha=2.509$ and $J=0$. This maximum growth rate value is very close to the one observed for the wind flow case. This can be seen by comparing Fig.\ \ref{fig:7}(a) with  Fig.\ \ref{fig:2}(a).  The stability boundaries are also very similar, some  small differences arise for higher wavenumbers. The phase speed contours depicted in Fig.\ \ref{fig:7}(b) are also quite similar to that of the wind flow case shown in Fig.\ \ref{fig:2}(b). Some small differences arise outside the stability boundaries. Therefore, it can be inferred that  consistent presence of wind or lack of it above the lake surface has no appreciable effect on the stability characteristics. Strong, persistent wind forcing event is only necessary for  setting up  the base state double circulation. Afterwards, instabilities can be initiated by small random disturbances, e.g.\ a gust of wind, but a consistent, mean wind flow is not necessary.  

\begin{figure}
  \centering{\includegraphics[width=5in]{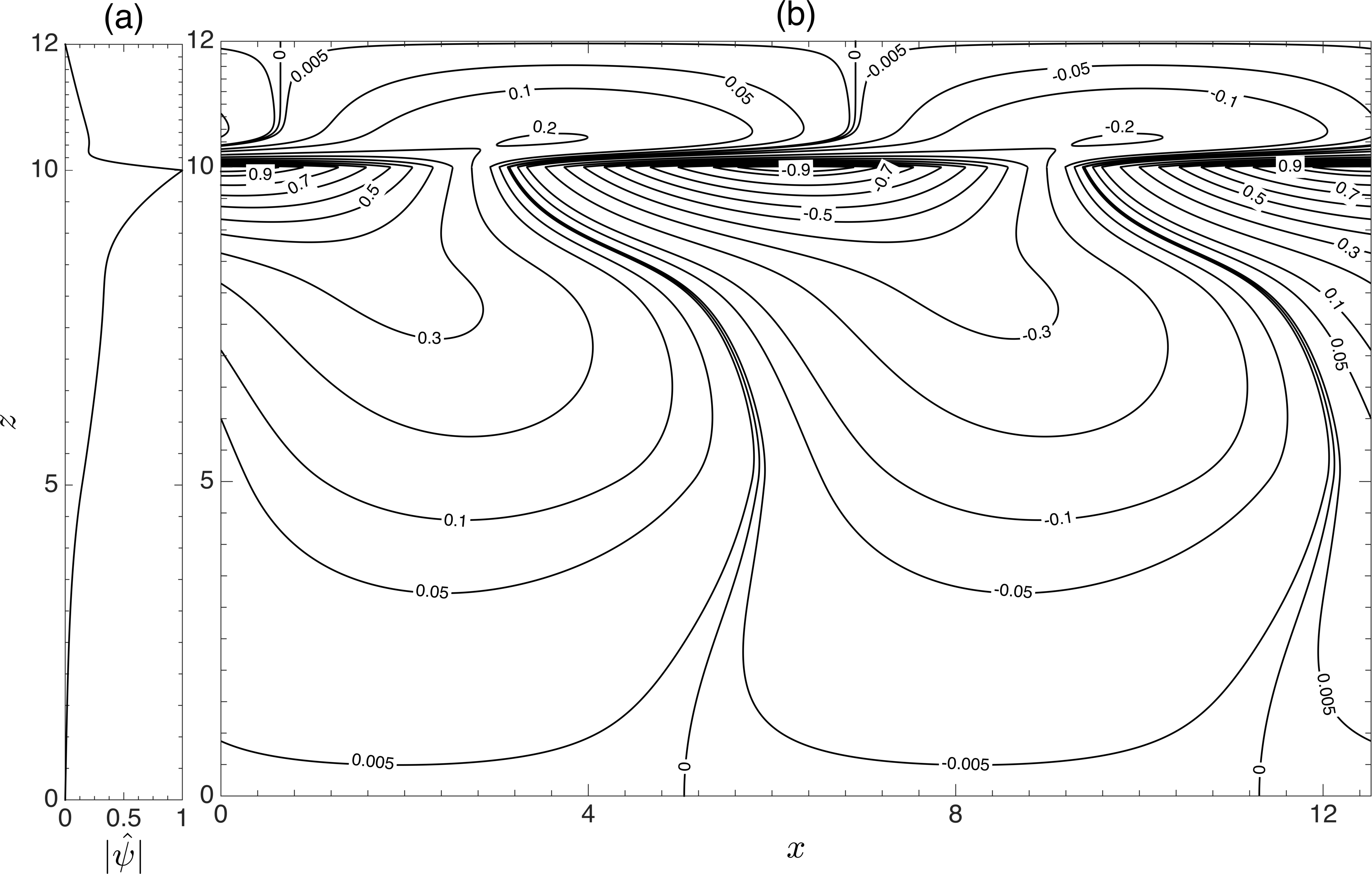}}
  \caption{Plots for non-Boussinesq perturbation stream function for the still air case: (a) Norm of  perturbation stream function eigenfunction versus $z$, and (b) contours of perturbation stream function $\tilde{\psi}$.}
\label{fig:8}
\end{figure}

In Fig.\ \ref{fig:8} we show the eigenfunctions of the perturbation stream function corresponding to the case of maximum growth rate ($\alpha=2.509$, $J=0$).   The reported eigenfunction  is normalized by its maximum value. Figure \ref{fig:8}(a) shows that $\hat{\psi}$ attains a maxima at the free-surface. The contours of $\tilde{\psi}$ for one wavelength of the disturbance are shown in Fig. \ref{fig:8}(b). Slight differences in $\hat{\psi}$ or $\tilde{\psi}$ between  quiescent air and wind flow case can be observed by comparing Fig.\ \ref{fig:3} and Fig.\ \ref{fig:8}.

\begin{figure}
  \centering{\includegraphics[width=5in]{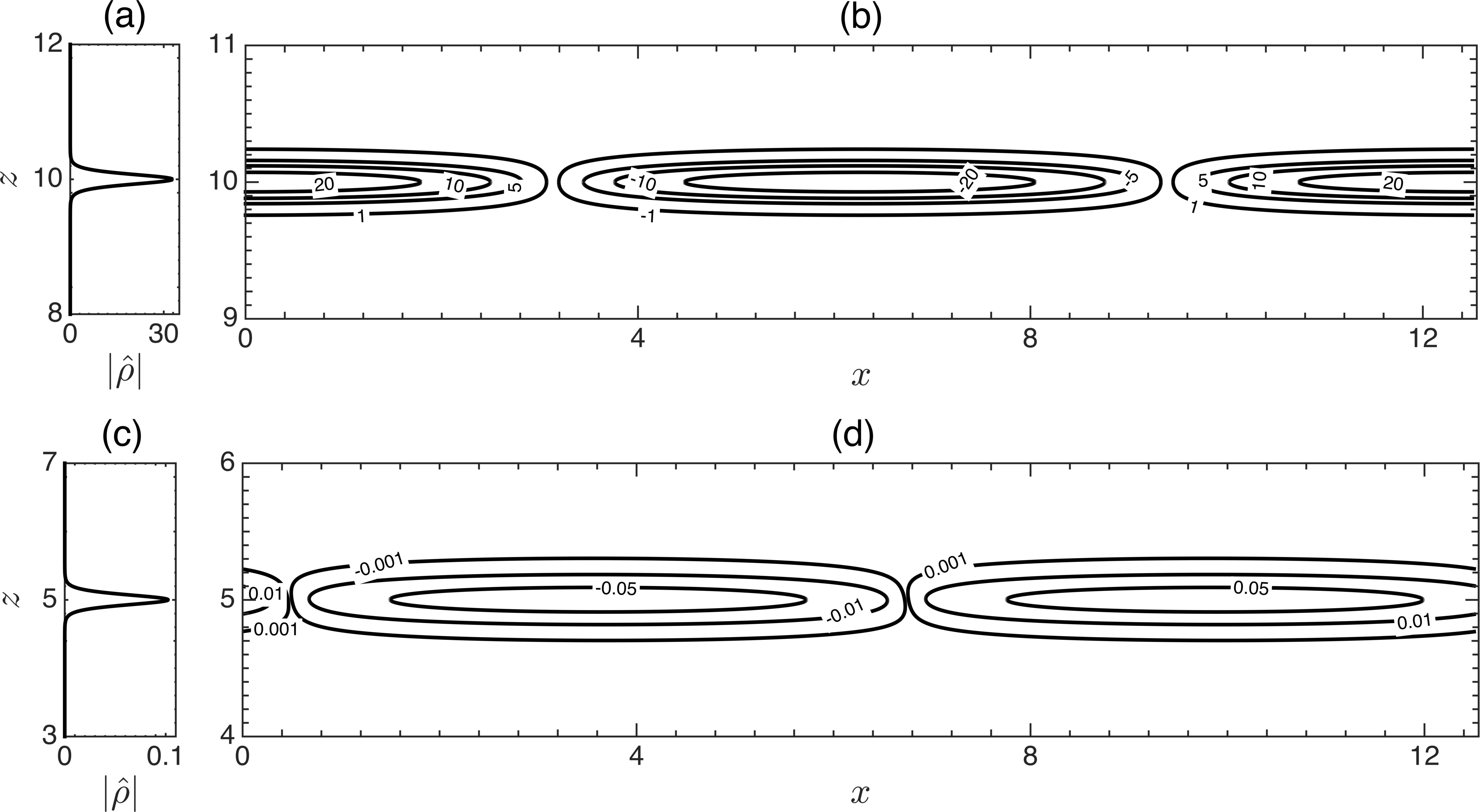}}
  \caption{Plots of non-Boussinesq  density perturbation for still air case. The perturbation density eigenfunctions are normalized by the maximum value of corresponding perturbation stream function eigenfunction. (a) Norm of perturbation density eigenfunction versus $z$ near the free-surface. (b) Contours of perturbation density, $\tilde{\rho}$, near the free-surface. (c) Norm of perturbation density eigenfunction versus $z$ near the pycnocline. (d) Contours of perturbation density, $\tilde{\rho}$, near the pycnocline.}
\label{fig:9}
\end{figure}

The perturbation density $\tilde{\rho}$, and the corresponding eigenfunction $\hat{\rho}$, are shown in Fig.\ \ref{fig:9}. The quantity $\hat{\rho}$ is normalized by the maximum value of the corresponding $\hat{\psi}$. 
The density contours in the vicinity of the free-surface are shown in Fig.\ \ref{fig:9}(b), while that near the pycnocline are shown in Fig.\ \ref{fig:9}(d). The perturbation density contours and the corresponding eigenfunctions are very similar to the ones obtained for the wind flow case as can be seen by comparing Fig. \ref{fig:4} with Fig. \ref{fig:9}. 
The physics of instability of the still air problem is almost exactly same as that of that of the wind flow problem described in \S \ref{sub:wind}.

\section{Stability of Boussinesq Continuous Profile}
\label{sec:4}
\begin{figure}
  \centering{\includegraphics[width=4in]{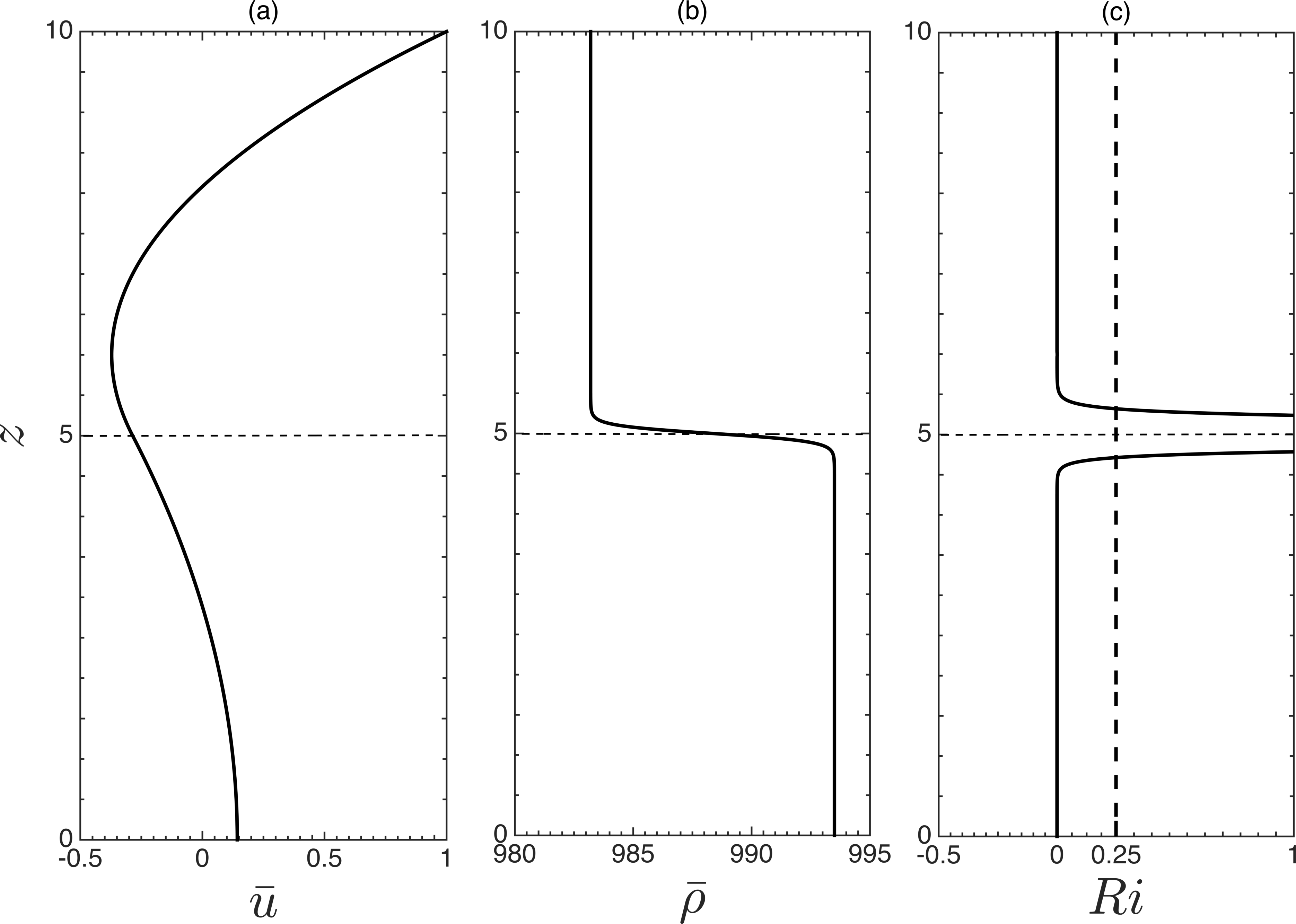}}
  \caption{Continuous velocity and density profiles for the Boussinesq case. (a) Velocity profile, (b) density profile depicting a Boussinesq jump, and (c) gradient Richardson number, $Ri$ versus $z$.  The dashed-vertical line corresponds to $Ri=0.25$. }
\label{fig:9.5}
\end{figure}

To contrast the non-Boussinesq stability analysis, a Boussinesq stability analysis is performed on the continuous velocity profile given by (\ref{eq:13}) and density profile given by (\ref{eq:15}). These profiles are plotted in  Fig.\ \ref{fig:9.5}. Boussinesq approximation can only work for small density differences (for example, that between colder and warmer water) but will fail to capture the effect of large density difference existing at the free-surface. Hence Boussinesq approximation  requires the free-surface to be replaced by rigid-lid.  A schematic of this case has been shown in  Fig.\ \ref{fig:0}(c).  Linear stability analysis is performed on the velocity and density profiles mentioned above. The variation of gradient Richardson number, $Ri$ with $z$ is shown in Fig. \ref{fig:9.5}(c). The quantity $Ri$ is defined as follows:
\[
Ri(z)=-\frac{g}{\rho_{0}}\frac{\frac{d\bar{\rho}}{dz}}{\left(\frac{d\bar{u}}{dz}\right)^{2}},
\]
where $\rho_{0}$ is the reference density (which we have taken to be the mean of the warmer and colder water). According to the Miles-Howard necessary condition of instability, a flow can become unstable if $Ri(z)<0.25$ somewhere inside the domain \cite[]{drazin2004hydrodynamic,schmid2012stability}. This criterion is satisfied in most of the domain, as can be observed in Fig.\ \ref{fig:9.5}(c), hence instability can be expected. For the presentation of the results discussed in this section  we have non-dimensionalized the wavenumber, phase speed and growth rates using the scales given in \S \ref{sub:wind}.

\begin{figure}
  \centering{\includegraphics[width=5in]{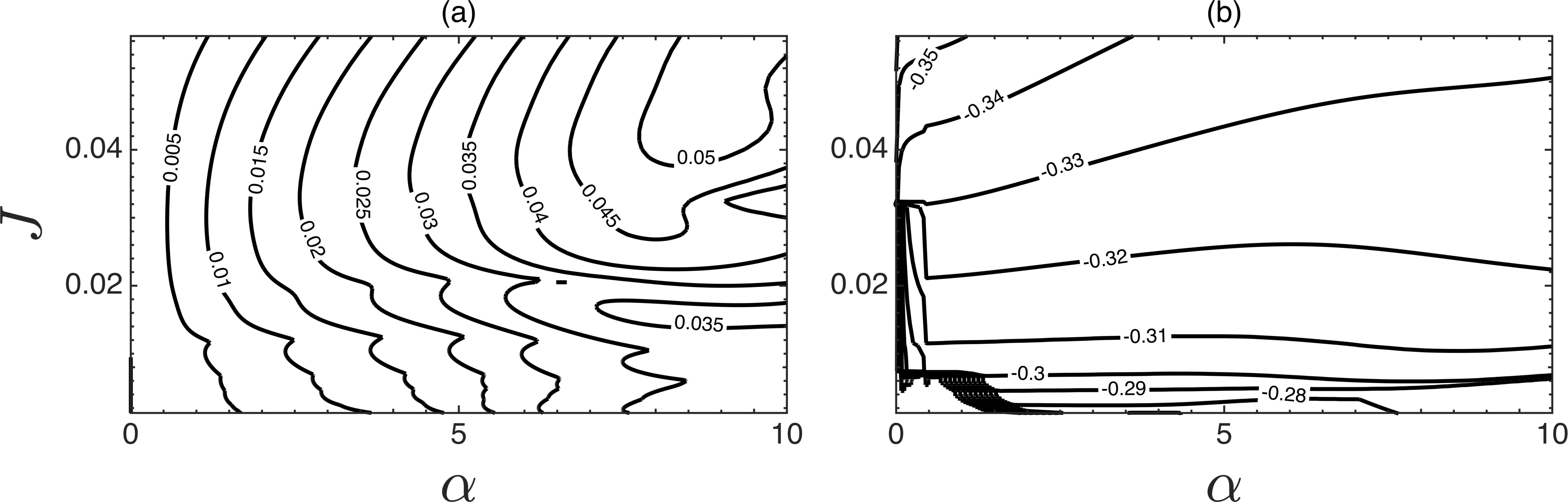}}
  \caption{Boussinesq growth rates and phase speeds: (a) growth rates and (b) phase speeds.}
\label{fig:10}
\end{figure}

In Fig.\ \ref{fig:10} we plot growth rates and phase speeds for the Boussinesq case. 
For $J=0$ (unstratified case), the growth rates are positive, not zero. This is because  Rayleigh's inflection point theorem \cite[]{lord1880stability}, which states that presence of an inflection point in the velocity profile is a necessary condition for instability, is satisfied. 
In this case an inflection point is present at $z=5$ (observe in Fig.\ \ref{fig:1}(b) that $d\bar{q}/dz$ changes sign at this location).  Fig.\ \ref{fig:10}(a) reveals a maximum growth rate of $0.0521$, which occurs at  $\alpha=9.3647$ and  $J=0.0567$. 
We find that the maximum Boussinesq growth rate is an order of magnitude lower than the non-Boussinesq ones observed in Fig.\ \ref{fig:7} or  Fig.\ \ref{fig:2}. Moreover the stability boundaries with and without Boussinesq approximation are very different. Differences are also observed in the phase speed plots. This can be seen by comparing Fig.\ \ref{fig:10}(b) with    Fig.\ \ref{fig:2}(b),  Fig.\ \ref{fig:2}(d) and Fig.\ \ref{fig:7}(b). Boussinesq case shows that the unstable waves have negative phase speeds, while they are positive for the non-Boussinesq  (both wind flow and still air) cases. We emphasize here that the comparisons between non-Boussinesq (with free-surface) and Boussinesq (with rigid-lid) results are the central points of this paper.  


\begin{figure}
	  \centering{\includegraphics[width=4.95in]{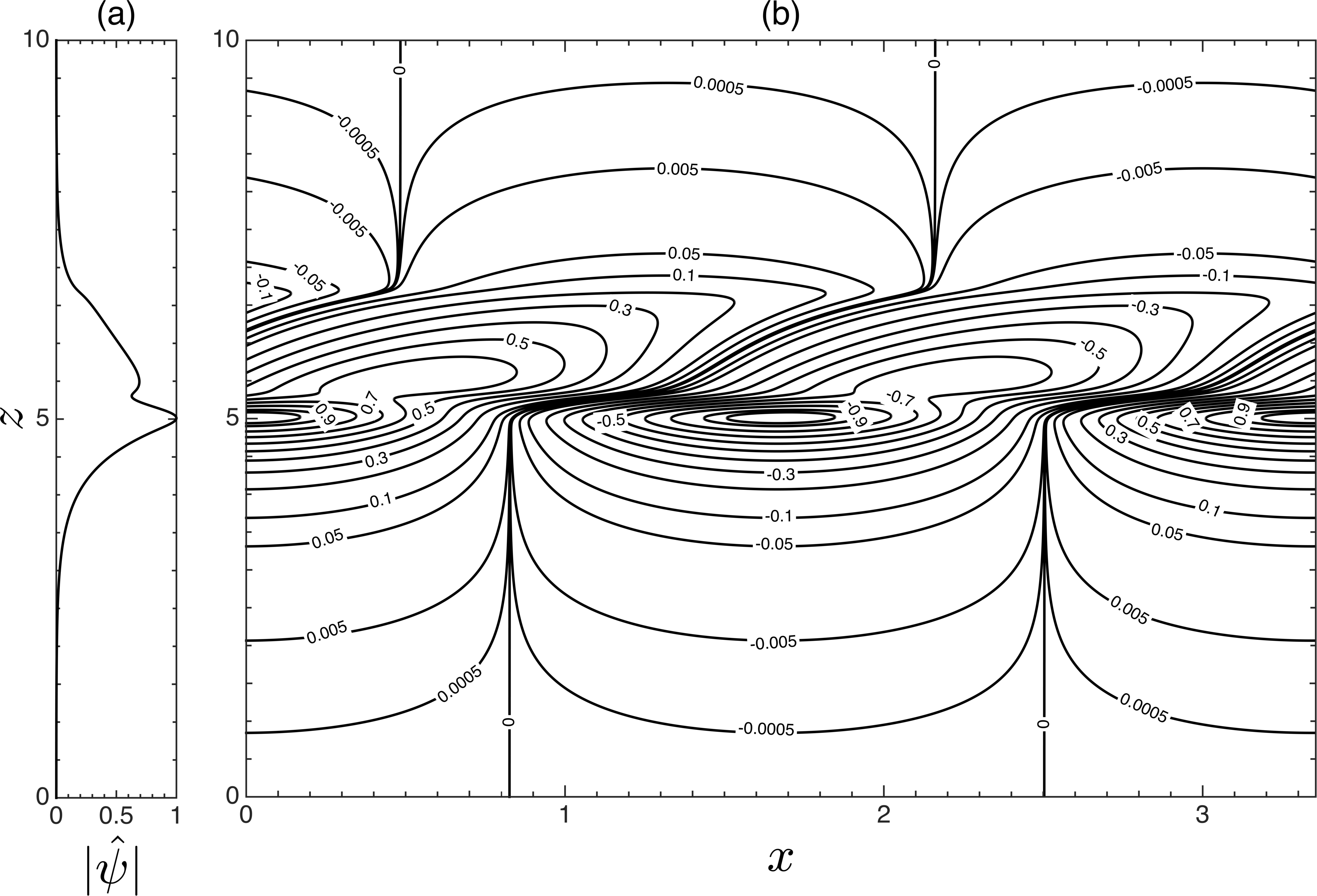}}
  \caption{Boussinesq perturbation stream function  plots: (a) Norm of  perturbation stream function eigenfunction versus $z$, and (b) contours of perturbation stream function $\tilde{\psi}$.}
\label{fig:11}
\end{figure}

\begin{figure}
\centering{\includegraphics[width=4.95in]{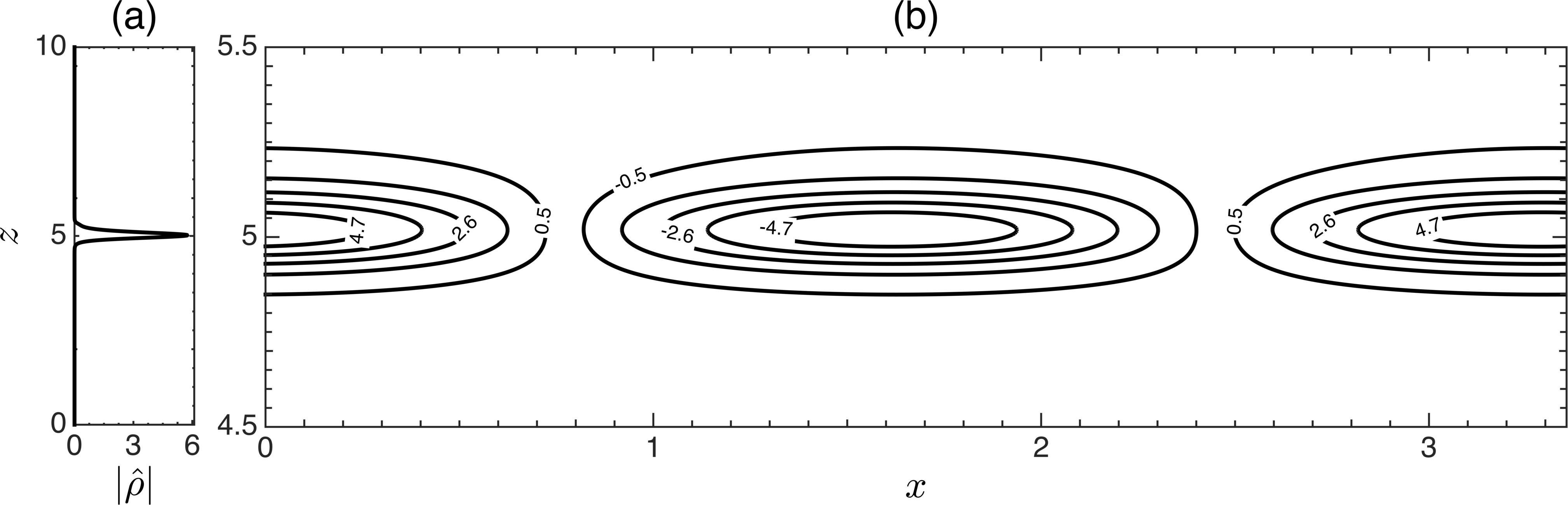}}\,\,\,
  \caption{Boussinesq  density perturbation plots. The perturbation density eigenfunction is normalized by the maximum  of the perturbation stream function eigenfunction. (a) Norm of perturbation density eigenfunction versus $z$. (b) Contours of perturbation density, $\tilde{\rho}$, near the pycnocline.}
\label{fig:12}
\end{figure}

In Fig.\ \ref{fig:11} we show the eigenfunctions of the perturbation stream function corresponding to the case of maximum growth rate ($\alpha=9.3647$ and  $J=0.0567$).   The reported eigenfunction  is normalized by its maximum value. Figure \ref{fig:11}(a) reveals that $\hat{\psi}$ attains a maxima at the pycnocline. The contours of $\tilde{\psi}$ for one wavelength of the disturbance are shown in Fig. \ref{fig:11}(b).
The eigenfunction $\hat{\rho}$ , and the corresponding perturbation density $\tilde{\rho}$, are respectively  shown in Fig.\ \ref{fig:12}(a) and Fig.\ \ref{fig:12}(b). The quantity $\hat{\rho}$ is normalized by the maximum value of the corresponding $\hat{\psi}$. 
The density contours in the vicinity of the pycnocline are shown in Fig.\ \ref{fig:12}(b). Both Fig.\ \ref{fig:12}(a) and Fig.\ \ref{fig:12}(b) reveal the existence of interfacial gravity waves at the pycnocline.

\subsection{Relationship Between 2D Instabilities and 3D Instabilities for Boussinesq Case}
\begin{figure}
  \centering{\includegraphics[width=4.2in]{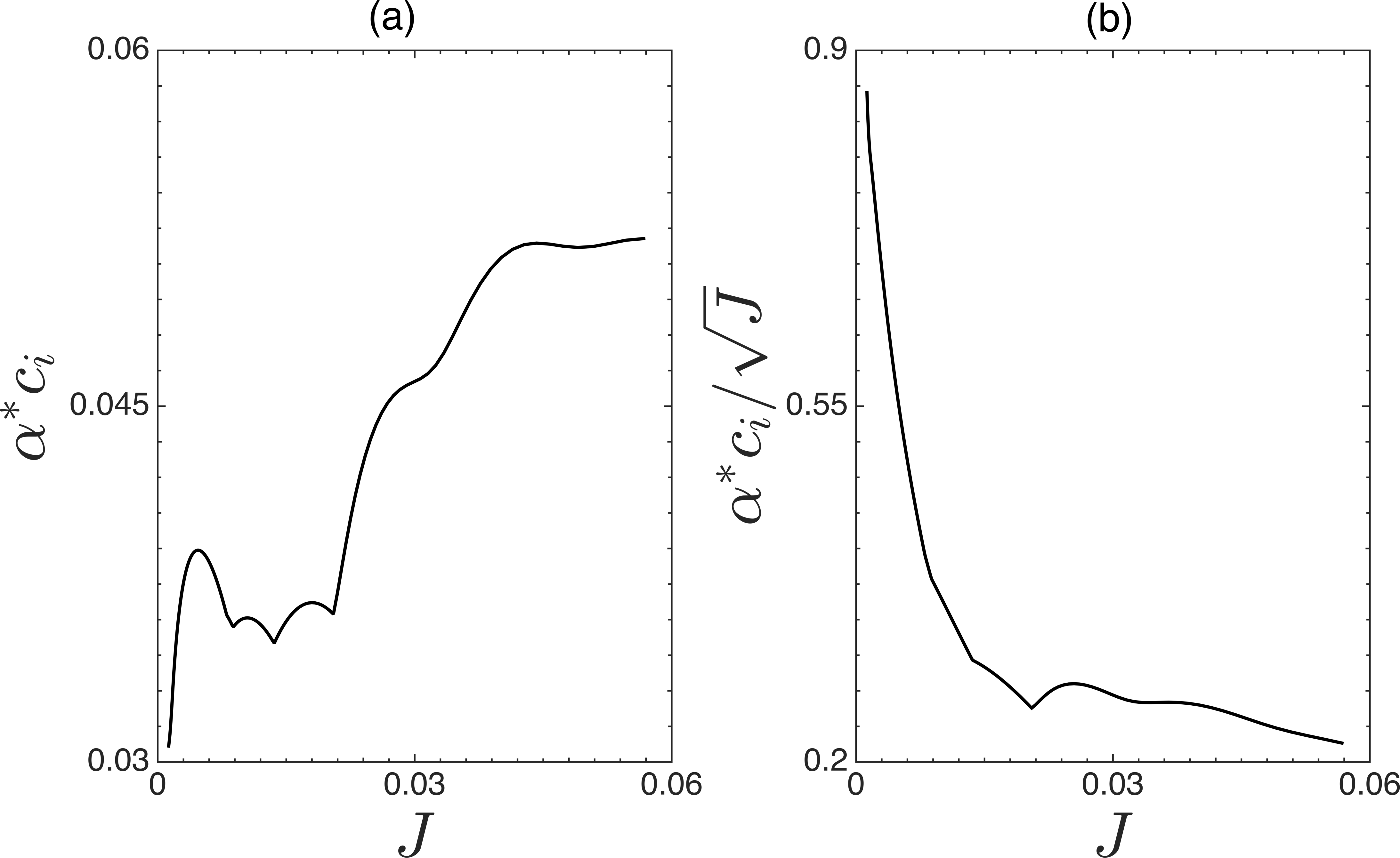}}
  \caption{(a) Variation of maximum growth rate, $\alpha^{*} c_{i}$ (evaluated at each $J$), versus $J$. (b) Plot of $\alpha^{*}c_{i}/\sqrt{J}$ versus $J$.}
\label{fig:13}
\end{figure}
The possibility of the occurrence of  three dimensional instabilities for the Boussinesq case is explored here. The procedure is same as that in \S \ref{sec:SP3D}. The variation of maximum growth rate, $\alpha^{*} c_{i}$, corresponding to each  value of $J$, is plotted in Fig.\ \ref{fig:13}(a). 
In Fig.\ \ref{fig:13}(b) we find that $\alpha^{*}c_{i}/\sqrt{J}$ as a function of $J$ is not a strictly monotonically decreasing function.  Hence it can be concluded that the instability might become three dimensional for some values of $J$. 

\section{Stability of Non-Boussinesq Broken-line Profiles}\label{sec:5}
\subsection{The Broken-line Profiles}
\begin{figure}
  \centering{\includegraphics[width=4.5in]{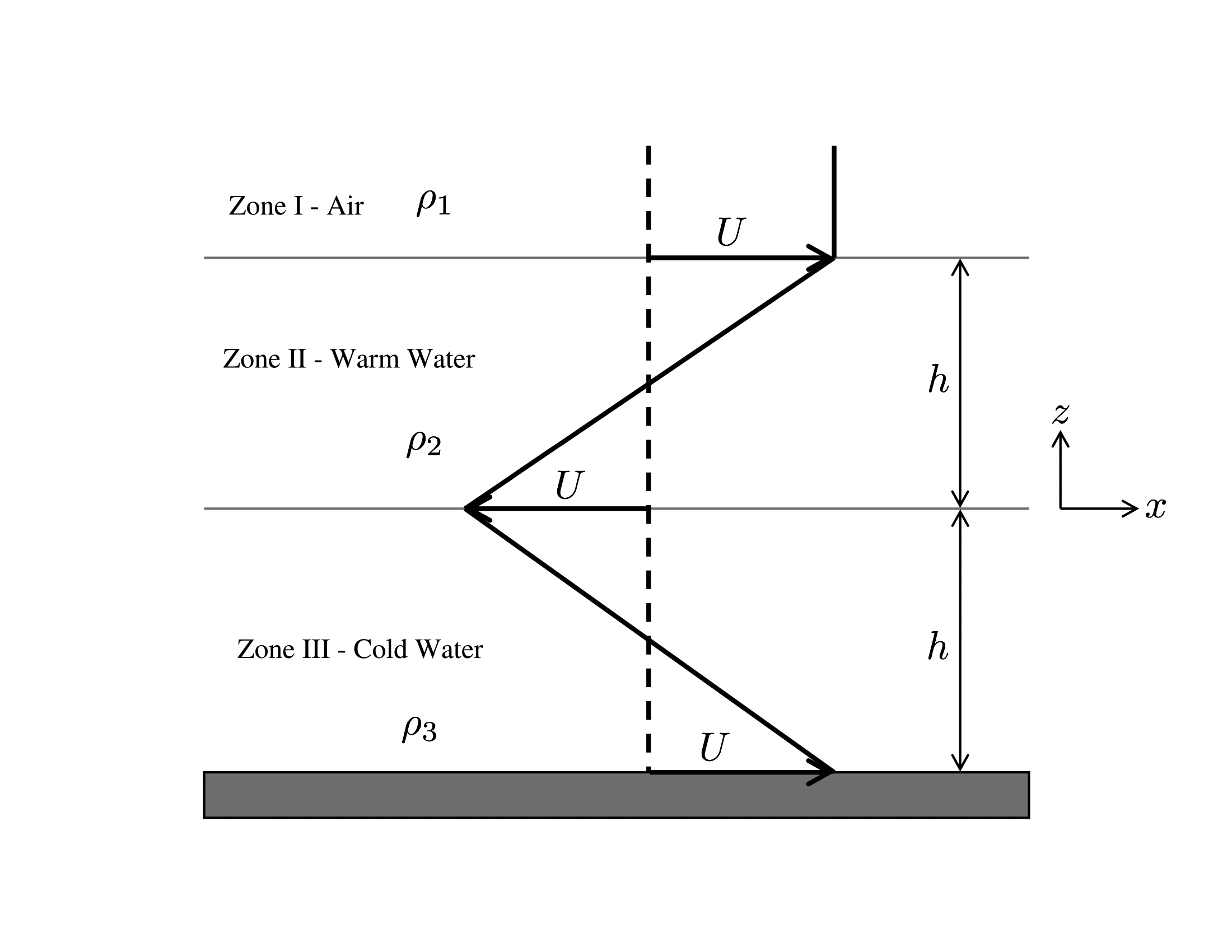}}
  \caption{ Schematic of the broken-line profile. The interface between zones I and II is the free-surface, and between zones II and III is the pycnocline. The gray shading denotes impermeable bottom boundary.}
\label{fig:14}
\end{figure}

 To complement the numerical stability analyses of the continuous profiles, we have also conducted an analytical study with simple broken-line profiles that capture the essence of the continuous, double circulation profile in Fig.\ \ref{fig:1}. The broken-line velocity and density profiles are given in the following equations, and  schematically shown in Fig.\ \ref{fig:14}. 

\begin{equation}\label{eq:17}
\bar{u}(z) = \left\{
        \begin{array}{cc}
        U & \quad z \geq h, \\ \\
            \frac{2 U}{h}z-U & \quad h \geq z \geq 0, \\ \\
            \frac{-2 U}{h}z-U  & \quad 0 \geq z \geq -h,
        \end{array}
    \right.
\end{equation}
\begin{equation}\label{eq:18}
\bar{\rho}(z) = \left\{
        \begin{array}{cc}
        \rho_{1} & \quad z > h, \\ \\
            \rho_{2} & h > z > 0, \\ \\
            \rho_{3}  & 0 > z > -h.
        \end{array}
    \right.
\end{equation}

\subsection{Governing Equations and Boundary Conditions}
In the inviscid and non-diffusive limit, both $\mu$, $\kappa \rightarrow 0$ . Therefore (\ref{eq:7}) and (\ref{eq:8}) can be simplified and combined to yield
\begin{equation}\label{eq:19}
\bar{\rho}^{\prime}\Big[(\bar{u}-c)\hat{w}^{\prime}- \bar{u}^{\prime} - \frac{g}{\bar{u}-c}\hat{w} \Big]  +  \bar{\rho}\Big[(\bar{u}-c)(\hat{w}^{\prime}-\alpha^{2}\hat{w}) - \bar{u}^{\prime\prime}\hat{w} \Big]=0.
\end{equation}
Equation (\ref{eq:19}) can be further simplified and converted into an equivalent form given by
\begin{equation}\label{eq:20}
\{\bar{\rho}[(\bar{u}-c)\hat{w}^{\prime} - \bar{u}^{\prime}\hat{w}]\}^{\prime} - \frac{\bar{\rho}^{\prime}g}{\bar{u}-c}\hat{w} - \bar{\rho}\alpha^{2}(\bar{u}-c)\hat{w} =0.
\end{equation}
The jump boundary condition across each interface can be derived by integrating (\ref{eq:20}) from $z_{i}+\Updelta z$ to $z_{i}-\Updelta z$ and letting the limit $\Updelta z \rightarrow 0$:
\begin{equation}\label{eq:21}
\llbracket \, \bar{\rho}[(\bar{u}-c)\hat{w}^{\prime}-\bar{u}^{\prime}\hat{w}-\frac{g\hat{w}}{\bar{u}-c}] \, \rrbracket=0.
\end{equation}
Here $z_{i}$ is the coordinate of the interface and $\llbracket . \rrbracket$ denotes the limits across the interface. The jump condition arises due to the continuity of pressure across the interfaces. Continuum hypothesis necessitates the kinematic condition, so that there are no gaps in the continuum. Kinematic condition across an  interface yields
\begin{equation}\label{eq:22}
\llbracket  \hat{w}  \rrbracket=0.
\end{equation}
Equation (\ref{eq:19}) along with the boundary conditions (\ref{eq:21}) and (\ref{eq:22}) are  separately solved in the three zones - air, warm water and cold water.

\subsection{Analytical Solution of Broken-line Profiles}
In each of the three zones mentioned above, (\ref{eq:19}) reduces to 
\begin{equation*}
\bar{\rho}(\bar{u}-c)[\hat{w}^{\prime\prime}-\alpha^{2}\hat{w}]=0.
\end{equation*}
Since we are only interested in the discrete eigenspectrum, we  assume $(\bar{u}-c)\neq 0$, hence
\begin{equation}\label{eq:23}
\hat{w}^{\prime\prime}-\alpha^{2}\hat{w}=0.
\end{equation}
On solving (\ref{eq:23}) in zone I (air) we get
\begin{equation}\label{eq:24}
\hat{w}=A_{1}\ee^{\alpha z} + B_{1}\ee^{-\alpha z}.
\end{equation}
On applying evanescence condition, i.e. $\hat{w}=0$ at $z\rightarrow \infty$ we get $A_{1}=0$. The eigenfunction in the air zone is given by
\begin{equation}\label{eq:25}
\hat{w}=B_{1}\ee^{-\alpha z}.
\end{equation}
Solving (\ref{eq:23}) in  zone II (warm water) produces
\begin{equation}\label{eq:26}
\hat{w}=A_{2}\ee^{\alpha z} + B_{2}\ee^{-\alpha z}.
\end{equation}
Similarly in zone III (cold water) we get
\begin{equation}\label{eq:27}
\hat{w}=A_{3}\ee^{\alpha z}+B_{3}\ee^{-\alpha z}.
\end{equation}
Impermeable boundary condition is applicable at the bottom boundary of the lake, which implies  $\hat{w}=0$ at $z\rightarrow -h$. This yields  $A_{3}=-B_{3}\ee^{2\alpha h}$. In the cold water zone the vertical velocity eigenfunction is given by
\begin{equation}\label{eq:28}
\hat{w}=-B_{3}\ee^{\alpha(z+2h)}+B_{3}\ee^{-\alpha z}.
\end{equation}
We simplify (\ref{eq:25}), (\ref{eq:26}) and (\ref{eq:28}) along with the jump conditions (\ref{eq:21}) and (\ref{eq:22}). Furthermore, we express $B_{1}$ and $B_{3}$ in terms of $A_{2}$ and $B_{2}$, and  use Boussinesq approximation at the pycnocline. Finally we obtain
\begin{equation}\label{eq:29}
\begin{bmatrix} 
\ee^{\alpha}\Big(\alpha(1-c) - \frac{J_{1}}{1-c}-2\Big)  & \ee^{-\alpha}\Big(-\alpha(1-c) - \frac{J_{1}}{1-c}-2\Big) \\
2\frac{\alpha(1+c)}{1-\ee^{2\alpha}}+4+\frac{J_{2}}{1+c} & 2\frac{\alpha(1+c)\ee^{2\alpha}}{1-\ee^{2\alpha}}+4+\frac{J_{2}}{1+c}  
\end{bmatrix}
\begin{bmatrix}
A_{2} \\
B_{2}
\end{bmatrix}
=0.
\end{equation}
Here we have non-dimensionalized the wavenumber, $\alpha$ by the half-width of the lake, $h$ while the phase speed, $c$ has been non-dimensionalized by the velocity scale, $U$. The scales used here are the same as the one used in \S \ref{sub:wind}.  
\begin{equation*}
\,\, J_{1}= (\frac{\rho_{2}-\rho_{1}}{\rho_{2}})\frac{gh}{U^{2}}, \,\, J_{2}= (\frac{\rho_{3}-\rho_{2}}{\rho_{2}})\frac{gh}{U^{2}}.
\end{equation*}
The quantities $\rho_{1}$, $\rho_{2}$  and $\rho_{3}$ are respectively the densities of air, warm water and cold water. The parameters $J_{1}$ and $J_{2}$ are respectively the  non-Boussinesq and Boussinesq bulk Richardson numbers. For non-trivial solutions (\ref{eq:29}) yields a quartic dispersion relation in $c$. The dispersion relation is given by 
\begin{equation}\label{eq:30}
P_{4}c^{4}+P_{2}c^{2}+P_{1}c+P_{0}=0, 
\end{equation}
where 
\begin{equation*}\label{eq:31}
P_{4}=2 \ee^{3\alpha}\alpha^{2}+2\ee^{-\alpha}\alpha^{2},\,\,\,\,\,\,\,\,\,\,\,\,\,\,\,\,\,\,\,\,\,\,\,\,\,\,\,\,\,\,\,\,\,\,\,\,\,\,\,\,\,\,\,\,\,\,\,\,\,\,\,\,\,\,\,\,\,\,\,\,\,\,\,\,\,\,\,\,\,\,\,\,\,\,\,\,\,\,\,\,\,\,\,\,\,\,\,\,\,\,\,\,\,\,\,\,\,\,\,\,\,\,\,\,\,\,\,\,\,\,\,\,\,\,\,\,\,\,\,\,\,\,\,\,\,\,\,\,\,\,\,\,\,\,\,\,\,\,\,\,\,\,\,\,\,\,\,\,\,\,\,\,\,\,\,\,\,\,
\end{equation*}
\begin{multline}
 P_{2}=(-2\ee^{3\alpha}+2\ee^{-\alpha})\alpha J1+ ( -\ee^{3\alpha}+\ee^{-\alpha}) \alpha J2+ ( -4\ee^{3\alpha}-4\ee^{-\alpha} ) \alpha^{2} \nonumber \\
 + (8\ee^{3\alpha}-8\ee^{-\alpha} ) \alpha -8\ee^{3\alpha}+16\ee^{\alpha}-8\ee^{-\alpha},
\end{multline}
\begin{multline}
P_{1}=[-8\ee^{\alpha}+(-4\ee^{3\alpha}+4\ee^{-\alpha})\alpha+4\ee^{3\alpha}+4\ee^{-\alpha}]J_{1}\nonumber \\
+[(2\ee^{3\alpha}-2\ee^{-\alpha})\alpha-2\ee^{3\alpha}-2\ee^{-\alpha}+4\ee^{\alpha}]J_{2},
\end{multline}
\begin{multline}
P_{0}=[-8\ee^{\alpha}+ (-2\ee^{3\alpha}+2\ee^{-\alpha})\alpha + 4\ee^{-\alpha} + 4\ee^{3\alpha}] J_{1}+ (\ee^{-\alpha}+\ee^{3\alpha}-2\ee^{\alpha}) J_{2}J_{1} \nonumber \\
+ [( -\ee^{3\alpha}+\ee^{-\alpha}) \alpha +2\ee^{3\alpha}+2\ee^{-\alpha} - 4\ee^{\alpha}] J_{2} + ( 2\ee^{3\alpha}+2\ee^{-\alpha}) \alpha^{2} \nonumber \\ 
+  ( -8\ee^{3\alpha}+8\ee^{-\alpha} ) \alpha -16\ee^{\alpha}+8\ee^{-\alpha}+8\ee^{3\alpha}.
\end{multline}

\begin{figure}
  \centering{\includegraphics[width=4.7in]{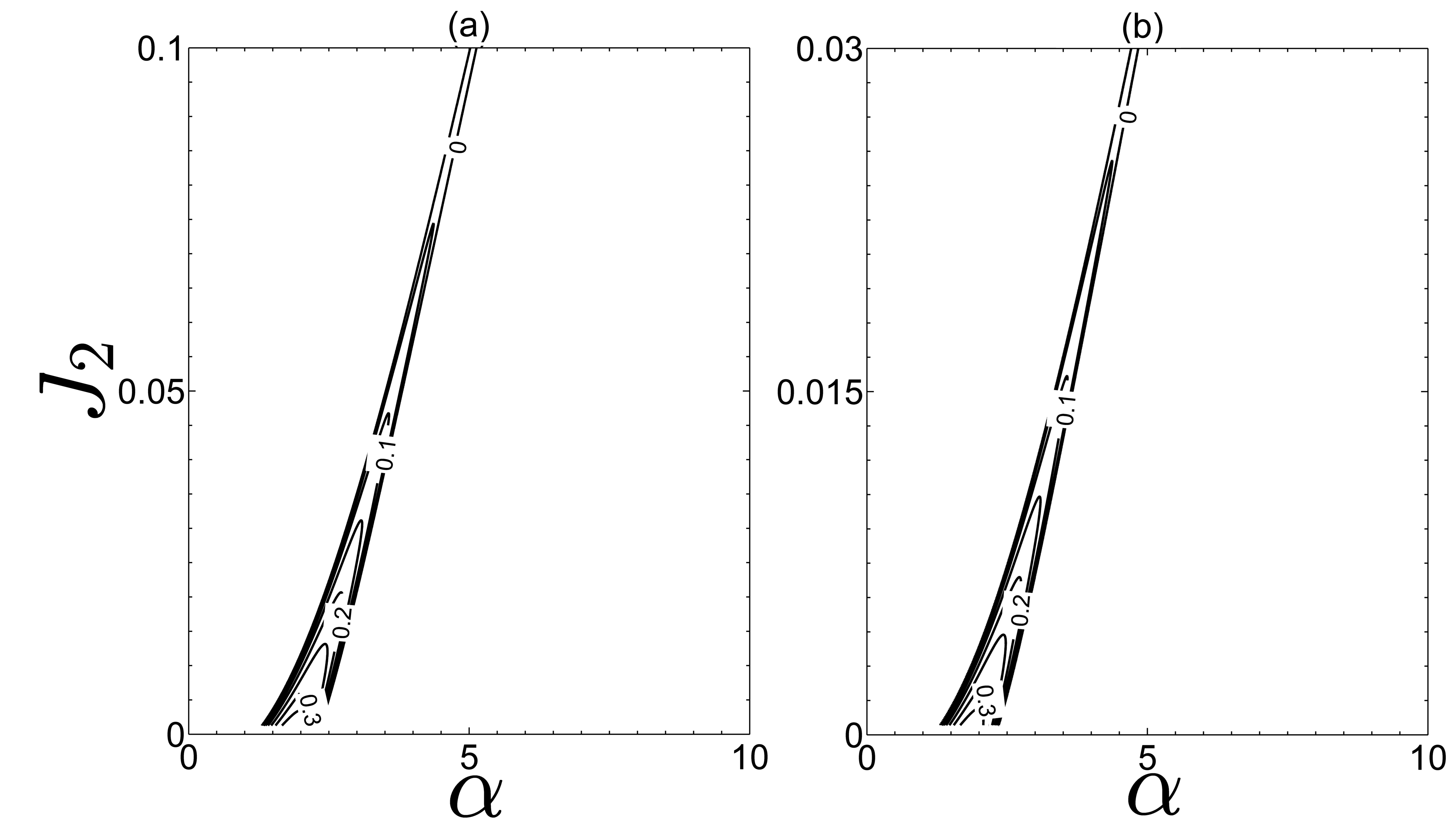}}
  \caption{Growth rate contour plots for the broken-line profiles.  (a) Growth rate contours for $At=0.01$ (b) Growth rate contours for $At=0.003$.}
\label{fig:15}
\end{figure}
We solve the quartic equation numerically using the MATLAB routine ``roots''. The non-dimensional growth rate plots for certain values of $At=J_{2}/J_{1}=(\rho_{3}-\rho_{2})/\rho_{2}$ are given in Fig.\ \ref{fig:15}. Growth rates for both Atwood numbers are the same but the stability boundaries are scaled versions of each other, the scaling factor being the ratio of the Atwood numbers. This scaling is due to the dependence of  $J_{2}$ on $At$ :  $J_{2}=At gh/U^{2}$. Such characteristics have also been observed in  Fig.\ \ref{fig:2}, highlighting the fact that the broken-line profile captures the essence of the continuous profile problem. 
The maximum growth rate of $0.3849$ occurs at a $J_{2}=0$ and $\alpha=1.8753$. The maximum growth rate as well as the wavenumber of the most unstable mode are comparable to the ones observed in numerical solution of continuous profiles. The broken-line profile underpins the physical reason behind this instability - it is due to the resonant interaction between the  surface vorticity-gravity wave and the interfacial vorticity-gravity wave, which are Doppler shifted by the base velocity. The wave interaction leading to this new kind of non-Boussinesq instability in lakes undergoing double circulation is markedly different from the Kelvin-Helmholtz instability, which is an interaction between two vorticity waves, and the Holmboe instability, that is an interaction between interfacial gravity wave and a vorticity wave \cite[]{caulfield1994multiple,baines1994mechanism,guha2014wave}. 

\section{Discussion and Conclusions}\label{sec:6}
We have performed  non-Boussinesq linear stability analyses of double circulation velocity profile that is possible in two-layered density stratified lakes. Two cases are considered - one assumes the air above the lake to be still, while the other takes wind flow into consideration. Non-Boussinesq analysis is required in order to capture the effect of the free-surface as well as the air above the lake on the stability of the flow inside the lake. Conventional linear stability analyses of stratified shear flows in the environment often makes use of Boussinesq approximation. Furthermore, such studies do not usually consider the finite extent of the vertical domain, and replace the free-surface by a rigid-lid.

 In \S \ref{sec:3} we have presented a brief account of the double circulation velocity profile that is possible in two-layered density stratified lakes. First we consider wind flow in the air above the lake, and then look into the quiescent air case. Stability analysis results for both cases are very comparable in terms of growth rates, phase speeds and stability boundaries, as can be seen from Fig.\ \ref{fig:2} and Fig.\ \ref{fig:7}. The maximum growth rate of around $0.48$ occurs for a non-dimensional wavenumber of around $2.5$ and bulk Richardson number of $0$. Density stratification, therefore, has a strictly stabilizing influence. We also infer that consistent presence of wind or lack of it above the lake surface has an insignificant effect on the stability characteristics. Strong and persistent wind forcing is necessary only for  setting up  the base state double circulation pattern. However, even a transient, small amplitude wind event (or any other random disturbances for that matter) may create perturbations, which can initiate this instability mechanism. The presence of surface gravity wave on the free-surface and the interfacial gravity wave on the pycnocline can be seen from Fig. \ref{fig:4}. Furthermore it has been shown that they can interact through the perturbation velocity field (see Fig. \ref{fig:3}) which is non-zero in the region between them, to produce exponential normal mode instability.

For comparison, we have also performed a conventional Boussinesq stability analysis of the double circulation profile  in \S \ref{sec:4}. Since Boussinesq analysis cannot account for the huge density difference occurring at the free-surface, it is replaced by a rigid-lid. The maximum Boussinesq growth rate is about $0.05$, which is an order of magnitude smaller than the maximum non-Boussinesq growth rate. Moreover the stability boundaries for the Boussinesq and non-Boussinesq cases are quite disparate. The phase speeds for the non-Boussinesq instability are positive while that for the Boussinesq instability are negative. The significant differences in the stability characteristics of the Boussinesq and non-Boussinesq cases arise because the latter is capable of considering waves on the free-surface. These surface waves can interact with the interfacial gravity waves at the pycnocline, and profoundly alter the stability characteristics.   
The Boussinesq as well as the non-Boussinesq instability can become three dimensional for some values of bulk Richardson number, $J$.

To complement our numerical stability study we have also conducted an analytical stability analysis of a simple broken-line profile in \S \ref{sec:5}. This profile captures the essential features of the double circulation profiles; see Fig. \ref{fig:14}. The maximum growth rate for the broken-line profiles is about $0.38$ and occurs for $J=0$ and $\alpha=1.87$. This is quite similar to the numerical stability of continuous non-Boussinesq profiles, where the maximum growth rate is about $0.48$ and occurs for $J=0$ and $\alpha=2.5$. Our analytical study with broken-line profile provides the physical reason behind this instability mechanism. It is due to the resonant interaction between the  surface vorticity-gravity wave (at the free-surface) and the interfacial vorticity-gravity wave (at the pycnocline), which are Doppler shifted by the base velocity.

In the current study we have presented evidence of the effect of surface waves on the stability of fluid below, in the case of two-layered density stratified lakes. Similarly  surface waves may also affect the stability of upper oceans (the ocean mixed layer), by interacting with the ocean pycnocline. Surface waves can potentially also affect the stability of estuaries and other exchange dominated flows. To the best of our knowledge all the previous studies have mostly modeled the free-surface as a rigid-lid and hence have essentially neglected the effect of surface waves on stability of the fluid below. 
 

\appendix
\section{Derivation of Non-Boussinesq Taylor-Goldstein Equation }\label{appA}
A full derivation of the non-Boussinesq viscous diffusive Taylor-Goldstein equation is given here.
We consider base states that are independent of time and are only dependent on the vertical coordinate, $z$. We define a base state given by $u=\bar{u}(z)$, $w=0$, $p=\bar{p}(z)$ and $\rho=\bar{\rho}(z)$, which also satisfies hydrostatic balance $d\bar{p}/dz=-\bar{\rho}g$.
Perturbation are added on top of the base state, yielding $u=\bar{u}(z)+\tilde{u}$, $w=\tilde{w}$, $p=\bar{p}(z)+\tilde{p}$, and $\rho=\bar{\rho}(z)+\tilde{\rho}$.  Here $\tilde{f}$ denotes perturbation quantities, where $f$ is the variable of interest. Substituting these in (\ref{eq:1}), (\ref{eq:2}), (\ref{eq:3}), (\ref{eq:6}) we get

\begin{equation}\label{eq:A7}
\frac{\partial\tilde{u}}{\partial x}+\frac{\partial\tilde{w}}{\partial z}=0,
\end{equation}
\begin{equation}\label{eq:A8}
(\tilde{\rho}+\bar{\rho}) \Big(\frac{\partial \tilde{u}}{\partial t}+\tilde{u}\frac{\partial \tilde{u}}{\partial x}+\bar{u}\frac{\partial \tilde{u}}{\partial x}+\tilde{w}\frac{\partial \tilde{u}}{\partial z}+\tilde{w}\frac{d\bar{u}}{dz}\Big)=-\frac{\partial \tilde{p}}{\partial x} +\mu\Big(\frac{\partial^{2}\tilde{u}}{\partial x^{2}}+\frac{\partial^{2}\tilde{u}}{\partial z^{2}}\Big),
\end{equation}
\begin{equation}\label{eq:A9}
(\tilde{\rho}+\bar{\rho}) \Big(\frac{\partial \tilde{w}}{\partial t}+\tilde{u}\frac{\partial \tilde{w}}{\partial x}+\bar{u}\frac{\partial \tilde{w}}{\partial x}+\tilde{w}\frac{\partial \tilde{w}}{\partial z}\Big)=-\frac{\partial \tilde{p}}{\partial z}-\frac{\partial \bar{p}}{\partial z}-\tilde{\rho}g-\bar{\rho}g+\mu\Big(\frac{\partial^{2}\tilde{w}}{\partial x^{2}}+\frac{\partial^{2}\tilde{w}}{\partial z^{2}}\Big),
\end{equation}
\begin{equation}\label{eq:A10}
\frac{\partial \tilde{\rho}}{\partial t}+\tilde{u}\frac{\partial \tilde{\rho}}{\partial x}+\bar{u}\frac{\partial \tilde{\rho}}{\partial x}+\tilde{w}\frac{\partial \tilde{\rho}}{\partial z}+\tilde{w}\frac{d\bar{\rho}}{dz}=\kappa\Big(\frac{\partial^{2}\tilde{\rho}}{\partial x^{2}}+\frac{\partial^{2}\tilde{\rho}}{\partial z^{2}}\Big).
\end{equation}
Assuming perturbations are infinitesimal, we  only retain the linear terms in (\ref{eq:A7})-(\ref{eq:A10}):

\begin{equation}\label{eq:A11}
\frac{\partial\tilde{u}}{\partial x}+\frac{\partial\tilde{w}}{\partial z}=0,
\end{equation}
\begin{equation}\label{eq:A12}
\bar{\rho} (\frac{\partial \tilde{u}}{\partial t}+\bar{u}\frac{\partial \tilde{u}}{\partial x}+\tilde{w}\frac{d\bar{u}}{dz})=-\frac{\partial \tilde{p}}{\partial x} +\mu\Big(\frac{\partial^{2}\tilde{u}}{\partial x^{2}}+\frac{\partial^{2}\tilde{u}}{\partial z^{2}}\Big),
\end{equation}
\begin{equation}\label{eq:A13}
\bar{\rho} (\frac{\partial \tilde{w}}{\partial t}+\bar{u}\frac{\partial \tilde{w}}{\partial x})=-\frac{\partial \tilde{p}}{\partial z}-\tilde{\rho}g+\mu\Big(\frac{\partial^{2}\tilde{w}}{\partial x^{2}}+\frac{\partial^{2}\tilde{w}}{\partial z^{2}}\Big),
\end{equation}
\begin{equation}\label{eq:A14}
\frac{\partial \tilde{\rho}}{\partial t}+\bar{u}\frac{\partial \tilde{\rho}}{\partial x}+\tilde{w}\frac{d\bar{\rho}}{dz}=\kappa\Big(\frac{\partial^{2}\tilde{\rho}}{\partial x^{2}}+\frac{\partial^{2}\tilde{\rho}}{\partial z^{2}}\Big).
\end{equation}
Following the normal mode stability theory outlined in  \cite{drazin2004hydrodynamic}, \cite{charru2011hydrodynamic} and \cite{schmid2012stability}, we introduce normal mode perturbations of the form $\tilde{f}(x,z,t)=\hat{f}(z)\ee^{\ii \alpha(x-ct)}$, and substitute in (\ref{eq:A11})-(\ref{eq:A14}):
\begin{equation}\label{eq:A15}
\ii\alpha \hat{u}+\hat{w}^{\prime}=0,
\end{equation}
\begin{equation}\label{eq:A16}
\bar{\rho}(\ii \alpha(\bar{u}-c) \hat{u}+\hat{w}\bar{u}^{\prime})=-\ii \alpha \hat{p} +\mu(\hat{u}^{\prime\prime}-\alpha^{2}\hat{u}),
\end{equation}
\begin{equation}\label{eq:A17}
\bar{\rho}(\ii \alpha(\bar{u}-c) \hat{w})=-\hat{p}^{\prime}-\hat{\rho}g+\mu(\hat{w}^{\prime\prime}-\alpha^{2}\hat{w}),
\end{equation}
\begin{equation}\label{eq:A18}
\ii \alpha(\bar{u}-c) \hat{\rho}+\hat{w}\bar{\rho}^{\prime}=\kappa(\hat{\rho}^{\prime\prime}-\alpha^{2}\hat{\rho}).
\end{equation}
Substituting $\hat{u}=\ii \hat{w}^{\prime}/\alpha$ we get
\begin{equation*}
\bar{\rho}[-(\bar{u}-c)\hat{w}^{\prime}+\bar{u}^{\prime}\hat{w}]=-\ii\alpha\hat{p}+\frac{\ii}{\alpha}\mu[\hat{w}^{\prime\prime\prime}-\alpha^{2}\hat{w}^{\prime}].
\end{equation*}
Ordinary derivative with respect to $z$ of the above equation yields 
\begin{equation}\label{eq:A19}
\bar{\rho}^{\prime}[-(\bar{u}-c)\hat{w}^{\prime}+\bar{u}^{\prime}\hat{w}]+\bar{\rho}[-\bar{u}^{\prime}\hat{w}^{\prime} -(\bar{u}-c)\hat{w}^{\prime\prime} +\bar{u}^{\prime\prime}\hat{w}+\bar{u}^{\prime}\hat{w}^{\prime}]=-\ii\alpha\hat{p}^{\prime}+\frac{\ii}{\alpha}\mu[\hat{w}^{\prime\prime\prime\prime}-\alpha^{2}\hat{w}^{\prime\prime}].
\end{equation}
From (\ref{eq:A17}) we express $\hat{p}^{\prime}$ in terms of other variables as follows:
\begin{equation*}
	\hat{p}^{\prime}=-\ii\bar{\rho}\hat{w} \alpha(\bar{u}-c) -\hat{\rho}g+\mu(\hat{w}^{\prime\prime}-\alpha^{2}\hat{w}).
\end{equation*}
Substituting  $\hat{p}^{\prime}$ from above equation into (\ref{eq:A19}) we obtain
\begin{equation}\label{eq:A20}
\bar{\rho}^{\prime}[-(\bar{u}-c)\hat{w}^{\prime}+\bar{u}^{\prime}\hat{w}]+\bar{\rho}[-(\bar{u}-c)\hat{w}^{\prime\prime}+\alpha^{2}(\bar{u}-c)\hat{w}+ \bar{u}^{\prime\prime}\hat{w}]=\ii\alpha\hat{\rho}g+\frac{\ii}{\alpha}\mu[\hat{w}^{\prime\prime\prime\prime}-2\alpha^{2}\hat{w}^{\prime\prime}+\alpha^{4}\hat{w}],
\end{equation}
\begin{equation}\label{eq:A21}
\ii \alpha(\bar{u}-c) \hat{\rho}+\hat{w}\bar{\rho}^{\prime}=\kappa(\hat{\rho}^{\prime\prime}-\alpha^{2}\hat{\rho}).
\end{equation}
The above two equations, (\ref{eq:A20}) and (\ref{eq:A21}), form the \emph{non-Boussinesq viscous diffusive Taylor-Goldstein equations}. In the inviscid limit $\mu \rightarrow 0$, and in the non-diffusive limit $\kappa \rightarrow 0$. These limiting conditions produce
\begin{equation}\label{eq:A22}
\bar{\rho}^{\prime}[-(\bar{u}-c)\hat{w}^{\prime}+\bar{u}^{\prime}\hat{w}]+\bar{\rho}[-(\bar{u}-c)\hat{w}^{\prime\prime}+\alpha^{2}(\bar{u}-c)\hat{w}+ \bar{u}^{\prime\prime}\hat{w}]=\ii\alpha\hat{\rho}g,
\end{equation}
\begin{equation}\label{eq:A23}
\ii \alpha(\bar{u}-c) \hat{\rho}+\hat{w}\bar{\rho}^{\prime}=0.
\end{equation}
Substitution of $\hat{\rho}$ from (\ref{eq:A23}) into (\ref{eq:A22}) yields
\begin{equation}\label{eq:A24}
\bar{\rho}^{\prime}\Big[(\bar{u}-c)\hat{w}^{\prime}- \bar{u}^{\prime}\hat{w} - \frac{g}{\bar{u}-c}\hat{w} \Big]  +  \bar{\rho}\Big[(\bar{u}-c)(\hat{w}^{\prime}-\alpha^{2}\hat{w}) - \bar{u}^{\prime\prime}\hat{w} \Big]=0.
\end{equation}
Equation (\ref{eq:A24}) is  the non-Boussinesq Taylor-Goldstein equation, which is similar to  the non-Boussinesq equation obtained by \cite{barros2011holmboe}, \cite{barros2014elementary} and \cite{carp2016}.
Equation (\ref{eq:A24}) can be converted into an equivalent form using exact differentials:
\begin{equation}\label{eq:A25}
\{\bar{\rho}[(\bar{u}-c)\hat{w}^{\prime} - \bar{u}^{\prime}\hat{w}]\}^{\prime} - \frac{\bar{\rho}^{\prime}g}{\bar{u}-c}\hat{w} - \bar{\rho}\alpha^{2}(\bar{u}-c)\hat{w} =0.
\end{equation}

\section{3D Perturbations in Non-Boussinesq Taylor-Goldstein Equuation}\label{appB}
We derive a non-Boussinesq Taylor-Goldstein equation that has a general three dimensional perturbations. We follow a procedure similar to the one given in \cite{haigh1995} and \cite{smyth1990three}.

Incompressible continuity equation gives
\begin{equation}\label{eq:B1}
\frac{\partial u}{\partial x}+\frac{\partial v}{\partial y}+\frac{\partial w}{\partial z}=0.
\end{equation}
Navier-stokes equations in the $x$, $y$ and $z$ directions are respectively given by
\begin{equation}\label{eq:B2}
\rho \Big(\frac{\partial u}{\partial t}+u\frac{\partial u}{\partial x}+v\frac{\partial u}{\partial y}+w\frac{\partial u}{\partial z}\Big)=-\frac{\partial p}{\partial x}+\mu\Big(\frac{\partial^{2}u}{\partial x^{2}}+\frac{\partial^{2}u}{\partial y^{2}}+\frac{\partial^{2}u}{\partial z^{2}}\Big),
\end{equation}

\begin{equation}\label{eq:B2.5}
\rho \Big(\frac{\partial v}{\partial t}+u\frac{\partial v}{\partial x}+v\frac{\partial v}{\partial y}+w\frac{\partial v}{\partial z}\Big)=-\frac{\partial p}{\partial y}+\mu\Big(\frac{\partial^{2}v}{\partial x^{2}}+\frac{\partial^{2}v}{\partial y^{2}}+\frac{\partial^{2}v}{\partial z^{2}}\Big),
\end{equation}
and 
\begin{equation}\label{eq:B3}
\rho \Big(\frac{\partial w}{\partial t}+u\frac{\partial w}{\partial x}+v\frac{\partial w}{\partial y}+w\frac{\partial w}{\partial z}\Big)=-\frac{\partial p}{\partial z}-\rho g+ \mu\Big(\frac{\partial^{2}w}{\partial x^{2}}+\frac{\partial^{2}w}{\partial y^{2}}+\frac{\partial^{2}w}{\partial z^{2}}\Big).
\end{equation}
We  consider an advection-diffusion equation for density given as follows:
\begin{equation}\label{eq:B4}
\frac{\partial \rho}{\partial t}+u\frac{\partial \rho}{\partial x}+v\frac{\partial \rho}{\partial y}+w\frac{\partial \rho}{\partial z}=\kappa\Big(\frac{\partial^{2}\rho}{\partial x^{2}}+\frac{\partial^{2}\rho}{\partial y^{2}}+\frac{\partial^{2}\rho}{\partial z^{2}}\Big).
\end{equation}
This holds only if density and stratifying agent are related through a linear equation of state.
We define a base state given by $u=\bar{u}(z)$, $w=0$, $p=\bar{p}(z)$ and $\rho=\bar{\rho}(z)$, which also satisfies hydrostatic balance $d\bar{p}/dz=-\bar{\rho}g$.
Perturbation are added on top of the base state, yielding $u=\bar{u}(z)+\tilde{u}$, $w=\tilde{w}$, $p=\bar{p}(z)+\tilde{p}$, and $\rho=\bar{\rho}(z)+\tilde{\rho}$. Substituting these into  (\ref{eq:B1})-(\ref{eq:B4}) we get

\begin{equation}\label{eq:B5}
\frac{\partial\tilde{u}}{\partial x}+\frac{\partial\tilde{v}}{\partial y}+\frac{\partial\tilde{w}}{\partial z}=0,
\end{equation}

\begin{equation}\label{eq:B6}
(\tilde{\rho}+\bar{\rho}) \Big(\frac{\partial \tilde{u}}{\partial t}+\tilde{u}\frac{\partial \tilde{u}}{\partial x}+\bar{u}\frac{\partial \tilde{u}}{\partial x}+\tilde{v}\frac{\partial \tilde{u}}{\partial y}+\tilde{w}\frac{\partial \tilde{u}}{\partial z}+\tilde{w}\frac{d\bar{u}}{dz}\Big)=-\frac{\partial \tilde{p}}{\partial x} +\mu\Big(\frac{\partial^{2}\tilde{u}}{\partial x^{2}}+\frac{\partial^{2}\tilde{u}}{\partial y^{2}}+\frac{\partial^{2}\tilde{u}}{\partial z^{2}}\Big),
\end{equation}

\begin{equation}\label{eq:B6.5}
(\tilde{\rho}+\bar{\rho}) \Big(\frac{\partial \tilde{v}}{\partial t}+\tilde{u}\frac{\partial \tilde{v}}{\partial x}+\bar{u}\frac{\partial \tilde{v}}{\partial x}+\tilde{v}\frac{\partial \tilde{v}}{\partial y}+\tilde{w}\frac{\partial \tilde{v}}{\partial z}\Big)=-\frac{\partial \tilde{p}}{\partial y}+\mu\Big(\frac{\partial^{2}\tilde{v}}{\partial x^{2}}+\frac{\partial^{2}\tilde{v}}{\partial y^{2}}+\frac{\partial^{2}\tilde{v}}{\partial z^{2}}\Big),
\end{equation}

\begin{equation}\label{eq:B7}
(\tilde{\rho}+\bar{\rho}) \Big(\frac{\partial \tilde{w}}{\partial t}+\tilde{u}\frac{\partial \tilde{w}}{\partial x}+\bar{u}\frac{\partial \tilde{w}}{\partial x}+\tilde{v}\frac{\partial \tilde{w}}{\partial y}+\tilde{w}\frac{\partial \tilde{w}}{\partial z}\Big)=-\frac{\partial \tilde{p}}{\partial z}-\frac{\partial \bar{p}}{\partial z}-\tilde{\rho}g-\bar{\rho}g+\mu\Big(\frac{\partial^{2}\tilde{w}}{\partial x^{2}}+\frac{\partial^{2}\tilde{w}}{\partial y^{2}}+\frac{\partial^{2}\tilde{w}}{\partial z^{2}}\Big),
\end{equation}

\begin{equation}\label{eq:B8}
\frac{\partial \tilde{\rho}}{\partial t}+\tilde{u}\frac{\partial \tilde{\rho}}{\partial x}+\bar{u}\frac{\partial \tilde{\rho}}{\partial x}+\tilde{v}\frac{\partial \tilde{\rho}}{\partial y}+\tilde{w}\frac{\partial \tilde{\rho}}{\partial z}+\tilde{w}\frac{d\bar{\rho}}{dz}=\kappa\Big(\frac{\partial^{2}\tilde{\rho}}{\partial x^{2}}+\frac{\partial^{2}\tilde{\rho}}{\partial y^{2}}+\frac{\partial^{2}\tilde{\rho}}{\partial z^{2}}\Big).
\end{equation}
The perturbations are assumed to be infinitesimal, therefore we only retain  the linear terms in (\ref{eq:B5})-(\ref{eq:B8}) to obtain

\begin{equation}\label{eq:B9}
\frac{\partial\tilde{u}}{\partial x}+\frac{\partial\tilde{v}}{\partial y}+\frac{\partial\tilde{w}}{\partial z}=0,
\end{equation}

\begin{equation}\label{eq:B10}
\bar{\rho} \Big(\frac{\partial \tilde{u}}{\partial t}+\bar{u}\frac{\partial \tilde{u}}{\partial x}+\tilde{w}\frac{d\bar{u}}{dz}\Big)=-\frac{\partial \tilde{p}}{\partial x} +\mu\Big(\frac{\partial^{2}\tilde{u}}{\partial x^{2}}+\frac{\partial^{2}\tilde{u}}{\partial y^{2}}+\frac{\partial^{2}\tilde{u}}{\partial z^{2}}\Big),
\end{equation}

\begin{equation}\label{eq:B10.5}
\bar{\rho} \Big(\frac{\partial \tilde{v}}{\partial t}+\bar{u}\frac{\partial \tilde{v}}{\partial x}\Big)=-\frac{\partial \tilde{p}}{\partial y}+\mu\Big(\frac{\partial^{2}\tilde{v}}{\partial x^{2}}+\frac{\partial^{2}\tilde{v}}{\partial y^{2}}+\frac{\partial^{2}\tilde{v}}{\partial z^{2}}\Big),
\end{equation}

\begin{equation}\label{eq:B11}
\bar{\rho} \Big(\frac{\partial \tilde{w}}{\partial t}+\bar{u}\frac{\partial \tilde{w}}{\partial x}\Big)=-\frac{\partial \tilde{p}}{\partial z}-\tilde{\rho}g+\mu\Big(\frac{\partial^{2}\tilde{w}}{\partial x^{2}}+\frac{\partial^{2}\tilde{w}}{\partial y^{2}}+\frac{\partial^{2}\tilde{w}}{\partial z^{2}}\Big),
\end{equation}

\begin{equation}\label{eq:B12}
\frac{\partial \tilde{\rho}}{\partial t}+\bar{u}\frac{\partial \tilde{\rho}}{\partial x}+\tilde{w}\frac{d\bar{\rho}}{dz}=\kappa\Big(\frac{\partial^{2}\tilde{\rho}}{\partial x^{2}}+\frac{\partial^{2}\tilde{\rho}}{\partial y^{2}}+\frac{\partial^{2}\tilde{\rho}}{\partial z^{2}}\Big).
\end{equation}

The general form of a three dimensional normal mode perturbations is $\tilde{f}(x,y,z,t)=\hat{f}(z)e^{\ii \alpha(x\cos\phi+y\sin\phi-ct)}$ as shown in \cite{white2006viscous}.  This represents a traveling wave whose amplitude varies with $z$ and which propagates at an angle $\phi$ to the $x$-axis. This form of disturbance is introduced for all the perturbation quantities. 

\begin{equation}\label{eq:B13}
\ii\alpha (\hat{u}\cos\phi+\hat{v}\sin\phi)+\hat{w}^{\prime}=0,
\end{equation}

\begin{equation}\label{eq:B14}
\bar{\rho}[\ii \alpha(\bar{u}\cos\phi-c) \hat{u}+\hat{w}\bar{u}^{\prime}]=-\ii \alpha\cos\phi \hat{p} +\mu(\hat{u}^{\prime\prime}-\alpha^{2}\hat{u}),
\end{equation}

\begin{equation}\label{eq:B14.5}
\bar{\rho}[\ii \alpha(\bar{u}\cos\phi-c) \hat{v}]=-\ii\alpha\sin\phi\hat{p}
+\mu(\hat{v}^{\prime\prime}-\alpha^{2}\hat{v}),
\end{equation}

\begin{equation}\label{eq:B15}
\bar{\rho}[\ii \alpha(\bar{u}\cos\phi-c) \hat{w}]=-\hat{p}^{\prime}-\hat{\rho}g+\mu(\hat{w}^{\prime\prime}-\alpha^{2}\hat{w}),
\end{equation}

\begin{equation}\label{eq:B16}
\ii \alpha(\bar{u}\cos\phi-c) \hat{\rho}+\hat{w}\bar{\rho}^{\prime}=\kappa(\hat{\rho}^{\prime\prime}-\alpha^{2}\hat{\rho}).
\end{equation}
 By  (\ref{eq:B14})$\times\cos\phi$ $+$ (\ref{eq:B14.5})$\times\sin\phi$ we obtain
\begin{multline*}
\bar{\rho}\ii\alpha(\bar{u}\cos\phi-c)[\hat{u}\cos\phi+\hat{v}\sin\phi]+\bar{\rho}\hat{w}\bar{u}^{\prime}\cos\phi \\=-\ii\alpha\hat{p}+\mu[\hat{u}^{\prime\prime}\cos\phi+\hat{v}^{\prime\prime}\sin\phi-\alpha^{2}(\hat{u}\cos\phi+\hat{v}\sin\phi)]. 
\end{multline*}
From (\ref{eq:B13}) we have $\hat{u}\cos\phi+\hat{v}\sin\phi=\ii\hat{w}^{\prime}/\alpha$ giving
\begin{equation}\label{eq:B16.5}
\bar{\rho}[-(\bar{u}\cos\phi-c)\hat{w}^{\prime}+\hat{w}\bar{u}^{\prime}\cos\phi]=-\ii\alpha\hat{p}+\frac{\ii\mu}{\alpha}[\hat{w}^{\prime\prime\prime}-\alpha^{2}\hat{w}^{\prime}]. 
\end{equation}
Taking ordinary derivative with respect to $z$ of the above equation we get, 
\begin{equation}\label{eq:B17}
\bar{\rho}^{\prime}[-(\bar{u}\cos\phi-c)\hat{w}^{\prime}+\hat{w}\bar{u}^{\prime}\cos\phi]+\bar{\rho}[-(\bar{u}\cos\phi-c)\hat{w}^{\prime\prime}+\hat{w}\bar{u}^{\prime\prime}\cos\phi]=-\ii\alpha\hat{p}^{\prime}+\frac{\ii\mu}{\alpha}[\hat{w}^{\prime\prime\prime\prime}-\alpha^{2}\hat{w}^{\prime\prime}]. 
\end{equation}
The quantity $\hat{p}^{\prime}$ in  (\ref{eq:B15}) is expressed in terms of other quantities as follows:
\begin{equation*}
	\hat{p}^{\prime}=-\ii\bar{\rho}\hat{w}\alpha(\bar{u}\cos\phi-c) -\hat{\rho}g+\mu(\hat{w}^{\prime\prime}-\alpha^{2}\hat{w}).
\end{equation*}
Substitution of $\hat{p}^{\prime}$ from the above equation into  (\ref{eq:B17}) yields
\begin{multline}\label{eq:B18}
\bar{\rho}^{\prime}[-(\bar{u}\cos\phi-c)\hat{w}^{\prime}+\bar{u}^{\prime}\hat{w}\cos\phi]+\bar{\rho}[-(\bar{u}\cos\phi-c)\hat{w}^{\prime\prime}+\alpha^{2}(\bar{u}\cos\phi-c)\hat{w}+ \bar{u}^{\prime\prime}\hat{w}\cos\phi]\\=\ii\alpha\hat{\rho}g+\frac{i}{\alpha}\mu[\hat{w}^{\prime\prime\prime\prime}-2\alpha^{2}\hat{w}^{\prime\prime}-\alpha^{4}\hat{w}],
\end{multline}

\begin{equation}\label{eq:B19}
\ii \alpha(\bar{u}\cos\phi-c) \hat{\rho}+\hat{w}\bar{\rho}^{\prime}=\kappa(\hat{\rho}^{\prime\prime}-\alpha^{2}\hat{\rho}).
\end{equation}
In the inviscid and non-diffusive limit both $\mu \rightarrow 0$ and $\kappa \rightarrow 0$, yielding
\begin{multline}\label{eq:B20}
\bar{\rho}^{\prime}[-(\bar{u}\cos\phi-c)\hat{w}^{\prime}+\bar{u}^{\prime}\hat{w}\cos\phi]\\+\bar{\rho}[-(\bar{u}\cos\phi-c)\hat{w}^{\prime\prime}+\alpha^{2}(\bar{u}\cos\phi-c)\hat{w}+ \bar{u}^{\prime\prime}\hat{w}\cos\phi]=\ii\alpha\hat{\rho}g,
\end{multline}
\begin{equation}\label{eq:B21}
\ii \alpha(\bar{u}\cos\phi-c) \hat{\rho}+\hat{w}\bar{\rho}^{\prime}=0.
\end{equation}
Substituting $\hat{\rho}$ from (\ref{eq:B21}) in (\ref{eq:B20}) we get
\begin{equation*}
\bar{\rho}^{\prime}\Big[(\bar{u}\cos\phi-c)\hat{w}^{\prime}- \bar{u}^{\prime}\hat{w}\cos\phi - \frac{g}{(\bar{u}\cos\phi-c)}\hat{w} \Big]  +  \bar{\rho}\Big[(\bar{u}\cos\phi-c)(\hat{w}^{\prime}-\alpha^{2}\hat{w}) - \bar{u}^{\prime\prime}\hat{w}\cos\phi \Big]=0.
\end{equation*}
Dividing throughout by $\cos\phi$ yields
\begin{equation}\label{eq:B22}
\bar{\rho}^{\prime}\Big[(\bar{u}-\frac{c}{\cos\phi})\hat{w}^{\prime}- \bar{u}^{\prime}\hat{w} - \frac{g}{\cos^{2}\phi(\bar{u}-\frac{c}{\cos\phi})}\hat{w} \Big]  +  \bar{\rho}\Big[(\bar{u}-\frac{c}{\cos\phi})(\hat{w}^{\prime}-\alpha^{2}\hat{w}) - \bar{u}^{\prime\prime}\hat{w} \Big]=0.
\end{equation}
Equation (\ref{eq:B22}) is the inviscid non-diffusive non-Boussinesq Taylor-Goldstein equation for three dimensional perturbations.

\bibliographystyle{jfm}
\bibliography{references}

\end{document}